\documentclass[%
 reprint,
superscriptaddress,
 amsmath,amssymb,
 aps,
prc,
]{revtex4-2}
\usepackage{overpic}
\usepackage{graphicx}
\usepackage{dcolumn}
\usepackage{bm}
\usepackage{xcolor}
\usepackage{hyperref}

\begin{document}

\date{\today}

\title{Search for the Double Poles of the Scattering  Matrix in Light Nuclei}

\def\GANIL{Grand Acc\'{e}l\'{e}rateur National d'Ions Lourds (GANIL), CEA/DSM - CNRS/IN2P3, BP 55027, F-14076 Caen Cedex, France}
 \def\CAS{CAS Laboratory of High precision Nuclear Spectroscopy, Institute of Modern Physics, Chinese Academy of Sciences, Lanzhou 730000, China}
 \def\School{School of Nuclear Science and Technology, University of Chinese Academy of Sciences, Beijing 100049, China}
\author{David Cardona Ochoa}
\affiliation{\GANIL}
\author{Marek P{\l}oszajczak}
\affiliation{\GANIL}
\author {Nicolas Michel}
\affiliation{\CAS}
\affiliation{\School}

\begin{abstract}
Exceptional points (EPs) are non-Hermitian degeneracies at which two eigenvalues and their eigenvectors coalesce, producing a defective Hamiltonian and a double pole of the $S$-matrix. Using the coupled-channel Gamow Shell Model with the $\ell=1$ spin-orbit strengths as control 
parameters, we locate and characterize EPs in $^6\text{Li}$, $^7\text{Li}$, $^7\text{Be}$, and $^8\text{Be}$, and analyze their imprint on energies, 
widths, phase rigidity, spectroscopic factors, elastic cross sections, survival probabilities, and spectral functions. The signatures of the EP in scattering and time-domain observables are found to be strongly channel-dependent, and threshold effects play a decisive role in 
determining the accessibility of the EP in parameter space.
\end{abstract}

\maketitle

\section{Introduction}
In non-Hermitian systems, an Exceptional Point (EP) is a point in parameter space 
where two or more complex eigenvalues and their corresponding eigenstates coalesce, 
forming a spectral singularity~\cite{Kato1995, Heiss_2012}. Given the ubiquity of 
open quantum systems, whose features are well described by non-Hermitian 
Hamiltonians~\cite{Rotter_2015, AshidaGongUeda2020}, EPs have gained theoretical 
and experimental interest in multiple areas of physics, such as 
atomic~\cite{PhysRevLett.99.173003} and molecular 
physics~\cite{PhysRevLett.103.123003}, optics and 
photonics~\cite{doi:10.1126/science.aar7709, Ozdemir2019}, quantum phase 
transitions~\cite{PhysRevA.43.4159, Heiss_2005}, sensing 
applications~\cite{PhysRevLett.112.203901}, among others~\cite{Heiss_2012}. 

Unlike degeneracies in Hermitian systems, diabolic points (DPs), where distinct 
eigenvectors always span the full Hilbert space, at an EP the Hamiltonian becomes 
defective: it can no longer be diagonalized and is instead reducible only to a 
Jordan block structure~\cite{Kato1995, Heiss2004, Moiseyev_2011}. This defectiveness 
is the cause of the properties that distinguish EPs from all other degeneracies: the 
non-orthogonality of the coalescing states, the branch-point topology of the 
eigenvalue surfaces, and the breakdown of standard perturbation theory in their 
vicinity~\cite{Heiss_2012, Berry2004, AshidaGongUeda2020}.

A direct consequence of the unique topology of EPs is the behavior of the geometric 
phase under parameter encirclement. At a Hermitian diabolic point, a closed loop 
returns each eigenvalue to itself and the eigenstate acquires a Berry phase of 
$\pi$, a sign flip with no state exchange~\cite{Berry1984, Simon1983}. At an EP, 
by contrast, the eigenvalues form branch points: a single loop exchanges them 
cyclically and maps each eigenstate onto the other, requiring two loops to restore 
the original state~\cite{Mailybaev2005, Berry2004}. Although the accumulated phase 
is also $\pi$ for a second-order EP, it now accompanies a state exchange rather 
than a sign flip. For higher-order EPs, the permutation structure becomes 
non-Abelian. Furthermore, because the Berry connection must be defined using the 
biorthogonal inner product, the encirclement is chiral: clockwise and 
counterclockwise loops yield different final states, a feature with no Hermitian 
analog~\cite{Doppler2016, Xu2016}.

This special topology also influences the spectral response of the system to 
perturbations, since the eigenvalues coalesce at a branch point of the Riemann 
surface, a perturbation $\epsilon$ splits them as $\Delta E \sim \epsilon^{1/n}$ 
for an EP of order $n$, far more sensitive than the linear splitting at ordinary 
degeneracies~\cite{Heiss_2012}. This hypersensitivity has motivated proposals for 
EP-based sensing~\cite{PhysRevLett.112.203901, Wiersig2014}, though the same 
sensitivity also amplifies noise through the Petermann factor, limiting the 
signal-to-noise benefit~\cite{Langbein2018}, and produces anomalous response 
functions such as non-Lorentzian lineshapes and mode-mixing in driven 
systems~\cite{Berry2003, Berry2004}.

The decay dynamics of a system in the time domain can also be affected by the 
presence of an EP. It has been shown that due to the nature of the time evolution 
operator and the existence of a lower bound in the energy spectrum, pure exponential 
decay cannot hold for the whole time interval~\cite{LFonda_1978, Peshkin_2014, 
PhysRevResearch.5.023183}. Three distinct regimes should be observed in the 
evolution of a decaying state: short-time behavior typically in the form of 
parabolic decay, giving rise to the quantum Zeno 
effect~\cite{MisraSudarshan1977, Itano1990}; an exponential region spanning several 
lifetimes; and non-exponential evolution at long times with an inverse power-law 
form. Due to the presence of an EP even this intermediate region is no longer purely 
exponential and is instead modified by another enveloping function as has been 
studied theoretically~\cite{10.1063/1.4983809, Jaiswal_2023} and 
experimentally~\cite{PhysRevE.75.027201}.

Additionally, the effects of proximity to an EP leave imprints on scattering 
observables. From this point of view, an EP manifests as a double pole in the 
S-matrix, which has two principal consequences. First, it drives a redistribution 
of decay widths among coupled resonances: one state broadens while others become 
trapped with anomalously long lifetimes~\cite{Rotter2009, EleuchRotter2013}, 
invalidating the standard Breit-Wigner picture. Second, the double-pole structure 
produces interference terms that cause major deviations from Lorentzian lineshapes, 
including split-peak behavior in single-channel 
scattering~\cite{PhysRevA.95.022117} accompanied by a $2\pi$ phase-shift 
jump~\cite{EHernandez_2000}, and sudden channel-dependent enhancements or 
suppressions in multichannel reactions~\cite{Heiss2010, Eleuch2018}.

Nuclear physics offers an ideal setting for exploring resonance phenomena, as it 
naturally bridges discrete quantum states and the scattering continuum through 
decays, captures, and virtual excitations, while spanning a broad range of resonance 
lifetimes and widths: from narrow near-threshold states to broad resonances. This 
diversity makes it a privileged ground for systematic studies of resonance formation, 
evolution, and mixing, as well as the bound-to-continuum transition, providing both 
the conceptual framework and experimental observables needed to test non-Hermitian 
and open quantum system approaches, including the shell model embedded in the 
continuum~\cite{OKOLOWICZ2003271}, which has previously been used to study 
exceptional points (EPs)~\cite{PhysRevC.80.034619}. The Gamow Shell Model 
(GSM)~\cite{michel:in2p3-00331689, Michel2021} is particularly well suited for 
this: by incorporating bound, resonant, and continuum single-particle states on 
equal footing through a complex-energy Berggren basis, it describes nuclear states 
across all binding regimes while preserving unitarity. This is especially significant 
given that the conventional tools of nuclear reaction theory rely on the assumption 
that resonances can be treated as isolated poles of the $S$-matrix, which breaks down 
near an EP where two poles merge into a double pole. In $R$-matrix 
theory~\cite{LaneThomas1958, DescouvemontBaye2010}, the internal Hamiltonian is 
Hermitian by construction and the $S$-matrix is expanded in simple, well-separated 
poles, making it unable to reproduce the double-pole structure at an EP. 
Single-level Breit-Wigner parametrizations go further in this assumption, 
approximating the cross section near each resonance as a Lorentzian arising from a 
single isolated pole, an approximation that fails qualitatively when two poles 
coalesce and interference terms produce the non-Lorentzian lineshapes discussed 
above. Distorted-wave approaches~\cite{Satchler1983} are perturbative in the 
residual interaction and implicitly assume well-defined, non-degenerate initial 
states; near an EP the eigenstates themselves coalesce and become linearly dependent, 
rendering the perturbative expansion ill-defined. The GSM, by contrast, treats the 
non-Hermitian continuum coupling exactly and can naturally accommodate the 
double-pole structure of the $S$-matrix, making it a necessary framework for 
correctly describing resonance spectra and reaction observables in regimes of 
strong continuum coupling~\cite{cardona, CardonaOchoa2026}.

This paper is organized as follows. In Section~\ref{sec:framework} we lay out the theoretical framework: the coupled-channel formulation of the Gamow Shell Model (Sec.~\ref{sec:gsm-cc}), the extraction of reaction cross sections from the asymptotic $S$-matrix (Sec.~\ref{sec:cross-section}), and the properties of exceptional points relevant to our analysis, including the divergence and mutual cancellation of biorthogonal expectation values and the double-pole structure of the $S$-matrix (Sec.~\ref{sec:ep}). Section~\ref{sec:results} applies this framework to three nuclear systems of increasing complexity: the $2^+$ doublet in $^6$Li, where a single decay channel is open at the EP (Sec.~\ref{sec:A6}); the $5/2^-$ doublet in the mirror nuclei $^7$Li and $^7$Be, where the EP in $^7$Be lies above two open channels and illustrates the channel-dependent fate of the split-peak signature (Sec.~\ref{sec:A7}); and the $2^+$ isospin doublet in $^8$Be, where the EP emerges only above the proton threshold through a channel-induced restructuring of the eigenstates (Sec.~\ref{sec:A8}). For each case we examine energy and width trajectories, phase rigidity, spectroscopic factors, elastic cross sections, and survival probabilities. Section~\ref{sec:conclusions} summarizes our findings and discusses their broader implications.

\section{Theoretical Framework} \label{sec:framework}
\subsection{Gamow shell model in coupled-channel representation} \label{sec:gsm-cc}

The Gamow Shell Model (GSM) provides a natural extension of the conventional nuclear shell model into the complex-energy plane, furnishing a unified treatment in which bound states, resonances, and continuum states appear on equal footing. Many-body states are expressed as superpositions of Slater determinants constructed from single-particle wave functions drawn from the Berggren ensemble~\cite{michel:in2p3-00331689, Michel2021}:
\begin{equation}
    |\Psi^{J}_\alpha\rangle = \sum_n C_n |\text{SD}_n\rangle, \hspace{2mm} |\text{SD}_n\rangle = |\phi_{i_1} \phi_{i_2} \dots \phi_{i_A}\rangle,
\end{equation}
where each single-particle state $|\phi_{i}\rangle$ may be a bound state, a resonant (Gamow) state, or a scattering state defined along a contour $L_+$ in the complex momentum plane. The completeness of this single-particle basis is ensured by the Berggren completeness relation~\cite{michel:in2p3-00331689}:
\begin{equation}
    \sum_{n \in \text{bound, res}} |\phi_n\rangle \langle \Tilde{\phi}_n| + \int_{L_+} |\phi(k)\rangle \langle \Tilde{\phi}(k)| \, dk = \hat{1} \ ,
\end{equation}
where $\langle \Tilde{\phi}(k)|$ is the time-reversed conjugate of $|\phi(k)\rangle$. Diagonalization of the many-body Hamiltonian $\hat{H}$ in this complex-energy basis, $\hat{H} |\Psi^{J}_\alpha\rangle = E_\alpha |\Psi^{J}_\alpha\rangle$,
produces complex eigenvalues $E_\alpha = \epsilon_\alpha - i\Gamma_\alpha/2$, whose real and imaginary parts encode the energy and decay width of the corresponding nuclear state, respectively. This structure allows the GSM to capture the physics at the interface of bound and continuum dynamics, providing a microscopic description of weakly bound nuclei and resonances alike.

For a resonant state with complex energy $E_\alpha = \epsilon_\alpha - i\Gamma_\alpha/2$, the time evolution of the wave function takes the form:
\begin{equation}
|\Psi_\alpha(t)\rangle = e^{-i\epsilon_\alpha t/\hbar} e^{-\Gamma_\alpha t / 2\hbar} |\Psi_\alpha\rangle,
\end{equation}
revealing an exponential decay with half-life $T_{1/2} = \hbar\ln2/\Gamma_\alpha$. Expectation values of observables $\hat{O}$ in this framework are generally complex:
\begin{equation}
\label{expv}
\langle \hat{O} \rangle_\alpha = \frac{\langle \Tilde{\Psi}_\alpha | \hat{O} | \Psi_\alpha \rangle}{\langle \Tilde{\Psi}_\alpha | \Psi_\alpha \rangle} = \text{Re}\,\langle \hat{O} \rangle_\alpha + i\,\text{Im}\,\langle \hat{O} \rangle_\alpha,
\end{equation}
where the real part retains the standard interpretation as the observable mean value, while the imaginary part reflects the intrinsic dynamical uncertainty associated with the finite lifetime of the resonant state.

Despite its rigor, the Slater determinant formulation of the GSM is not directly amenable to the computation of reaction observables, since scattering boundary conditions and the asymptotic behavior of reaction channels cannot be imposed in that representation. To access quantities such as cross sections and phase shifts, one turns to the coupled-channels representation of the GSM (GSM-CC)~\cite{Michel2021,PhysRevC.89.034624, Wang2017}. In this formulation, the $A$-body wave function is expanded over a set of reaction channels:
\begin{equation}
    |\Psi^J\rangle = \sum_c \int_0^\infty |(c,r)^J\rangle \frac{u_c^{J}(r)}{r}r^2 dr,
    \label{react_channels}
\end{equation}
where $r$ denotes the separation between the centers of mass of the target and projectile clusters, and the magnetic quantum number $M$ is suppressed since all observables are independent of it. The radial amplitude $u_c^{J}(r)$ governs the relative motion in channel $c$ and is obtained by solving the GSM-CC equations for each value of the total angular momentum $J$. Although in principle the channel sum should include all possible binary, ternary, and higher-order partitions, in practice it is truncated; most GSM-CC applications have been restricted to binary channels, with only a few exceptions~\cite{Wang2017,PhysRevC.108.044616}. Each binary channel state $|(c,r)^J\rangle$ is defined through an antisymmetrized tensor product of the target $|\Psi_T^{J_T}\rangle$ and projectile $|\Psi_P^{J_P}\rangle$ states:
\begin{equation}
 |(c,r)^J\rangle = \hat{\mathcal{A}}\left[ |\Psi_T^{J_T}\rangle \otimes |\Psi_P^{J_P}\rangle\right]^{J},
\end{equation}
where $c$ labels the mass partition and its associated quantum numbers, and $J$ is obtained by coupling $J_T$ and $J_P$. Projecting the Schrödinger equation $\hat{H}|\Psi^J\rangle = E |\Psi^J\rangle$ onto the channel basis yields the coupled-channel equations:
\begin{equation}
    \sum_c \int_0^\infty r^2 \left(H_{cc'}(r, r') - E N_{cc'} (r,r')\right) \frac{u_c^J(r)}{r} =0, \label{schreq}
\end{equation}
with the Hamiltonian and norm kernels defined as:
\begin{equation}
    H_{cc'}(r,r') = \langle(c,r)^J|\hat{H}|(c',r')^J\rangle,
\end{equation}
\begin{equation}
    N_{cc'}(r,r') = \langle(c,r)^J|(c',r')^J\rangle.
\end{equation}
Exploiting the short-range nature of the internucleon interactions, the Hamiltonian is decomposed as:
\begin{equation}
    \hat{H}=\hat{H}_T + \hat{H}_P + \hat{H}_{TP},
\end{equation}
where $\hat{H}_T$ and $\hat{H}_P$ are the intrinsic Hamiltonians of the target and projectile, respectively. The projectile Hamiltonian is further separated as $\hat{H}_{P}=\hat{H}_{\rm int} + \hat{H}_{\rm CM}$, distinguishing its intrinsic dynamics from the center-of-mass motion. The residual cluster–cluster interaction is then given by $\hat{H}_{TP} = \hat{H}-\hat{H}_{T}-\hat{H}_{P}$, where $\hat{H}$ is the full shell-model Hamiltonian.

\subsection{Cross Section} \label{sec:cross-section}

The solution of Eq.~\eqref{schreq} yields the radial amplitudes $u_c^J(r)$, whose 
asymptotic behavior encodes all reaction information. Due to the non-locality of the 
GSM-CC equations when channel coupling is strong, direct integration in coordinate 
space via iterative methods such as the equivalent potential 
method~\cite{PhysRevC.89.034624} is not feasible. Instead, a numerical technique 
exploiting the completeness of the Berggren basis has been 
developed~\cite{Michel2021, PhysRevC.99.044606}, whereby Eq.~\eqref{schreq} is 
transformed into either a matrix diagonalization problem for bound and resonance 
states, or a linear system for scattering states.

The presence of the norm kernel $N_{cc'}(r,r')$, arising from the non-orthogonality 
of channel states constructed from Berggren basis wave functions, requires a 
regularization procedure before asymptotic matching can be 
performed~\cite{michel:in2p3-00331689, Michel2021}. This is achieved by 
diagonalizing the overlap matrix in the asymptotic region, thereby transforming the 
coupled-channel equations into a standard reaction form with orthogonal channels.

In the asymptotic region where the cluster--cluster interaction vanishes, the 
regularized channel amplitudes are matched to linear combinations of incoming and 
outgoing Coulomb--Hankel functions:
\begin{equation}
    u_c^J(r) \xrightarrow{r \to \infty} \frac{i}{2} \left[ 
    \delta_{cc'} H^{(-)}_{\ell_c}(\eta_c, k_c r) 
    - S_{cc'}^J H^{(+)}_{\ell_c}(\eta_c, k_c r) \right],
\end{equation}
where $e$ denotes the entrance channel, $H^{(\pm)}_\ell$ are the incoming/outgoing 
Coulomb--Hankel functions, $\eta_c = Z_T Z_P e^2 \mu_c / (\hbar^2 k_c)$ is the 
Sommerfeld parameter, and $k_c$ is the relative wave number in channel $c$. This 
matching procedure defines the partial-wave $S$-matrix elements $S_{cc'}^J$, from 
which all scattering observables can be extracted.

The differential cross section for the transition from entrance channel $c$ to exit 
channel $c'$ is obtained from the partial-wave expansion of the scattering amplitude:
\begin{equation}
    \frac{d\sigma_{c \to c'}}{d\Omega}(\theta) = \frac{1}{k_c^2} 
    \left| \sum_J (2J+1) \frac{S_{cc'}^J - \delta_{cc'}}{2i} P_J(\cos\theta) \right|^2,
    \label{eq:dxsec}
\end{equation}
where $P_J(\cos\theta)$ are the Legendre polynomials and the sum runs over all 
contributing total angular momenta. Integration over the solid angle yields the 
angle-integrated cross section. In the presence of Coulomb interactions, the nuclear 
cross section is extracted by subtracting the pure Coulomb contribution, and the 
interference between Coulomb and nuclear amplitudes must be properly accounted for.

The GSM-CC formalism constitutes a complex-energy extension of the resonating 
group method (RGM)~\cite{Michel2021}, with the key distinction that target and 
projectile states are GSM eigenstates rather than conventional shell-model states. 
This unified approach has been successfully applied to describe proton elastic and 
inelastic scattering on $^{18}$Ne~\cite{PhysRevC.89.034624}, deuteron--$\alpha$ 
elastic scattering in $^6$Li~\cite{PhysRevC.99.044606}, $^3$H and $^3$He elastic 
scattering on $^4$He in $^7$Li and $^7$Be~\cite{PhysRevC.108.044616}, radiative 
capture reactions such as $^8$Li$(n,\gamma)^9$Li~\cite{Dong2022}, and more recently 
$^{17}$Ne$(p,p)$ scattering~\cite{Chen2025}. In all cases, the same Hamiltonian 
describes both the structure of the nuclear states and the reaction cross sections, 
providing a truly unified framework for nuclear structure and reactions.

\subsection{Survival probability and spectral function} \label{sec:surv}

Beyond static reaction observables, the time evolution of a prepared state 
provides complementary information on the structure of resonances and the way 
they couple to the continuum. A natural object for this purpose is the 
\textit{spectral function} $\rho(E)$, defined as the projection of a many-body 
state $|\Psi\rangle$ onto the free particle eigenbasis $|\Psi_{E;c_n}\rangle$,
\begin{equation}
    \rho(E) = \sum_n|\langle \Psi_{E;c_n} | \Psi \rangle|^2,
\end{equation}
which characterizes the energy profile of the state and encodes both its 
localization in energy and its decay characteristics.

From the spectral function one obtains the \textit{survival probability} 
$\mathcal{S}(t)$, which measures the probability that the system, prepared in 
the state $|\Psi\rangle$, remains in this initial state after a time $t$. It is 
given by the squared modulus of the survival amplitude,
\begin{equation}
    A(t) = \langle \Psi | e^{-i\hat{H}t/\hbar} | \Psi \rangle 
         = \int dE \, \rho(E) \, e^{-iEt/\hbar},
\end{equation}
so that
\begin{equation}
    \mathcal{S}(t) = |A(t)|^2 
                   = \left| \int dE \, \rho(E) \, e^{-iEt/\hbar} \right|^2.
\end{equation}
The survival probability is therefore the Fourier transform of the spectral 
function, and the existence of a lower bound in the energy spectrum together 
with the deviations of $\rho(E)$ from a pure Lorentzian shape directly control 
the departures from purely exponential decay~\cite{LFonda_1978, Peshkin_2014, 
PhysRevResearch.5.023183}. As a consequence, three regimes generically appear in 
$\mathcal{S}(t)$: a short-time region in which $\mathcal{S}(t)$ is parabolic, 
giving rise to the quantum Zeno effect~\cite{MisraSudarshan1977, Itano1990}; an 
intermediate, approximately exponential regime governed by the resonance width; 
and an inverse power-law tail at long times of the form 
$\mathcal{S}(t)\sim t^{-(2l+3)}$, set by the threshold behavior of $\rho(E)$ 
near the lower edge of the spectrum~\cite{LFonda_1978, Peshkin_2014, 
PhysRevA.84.013419}.

\subsection{Exceptional Points} \label{sec:ep}

An EP occurs when, for a Hamiltonian $H(\lambda)$ represented by a complex
symmetric matrix, one can find $\lambda = \lambda_{\mathrm{EP}}$ such that
two eigenvalues and their eigenfunctions simultaneously coalesce,
\begin{equation}
    E_1(\lambda_{\mathrm{EP}}) = E_2(\lambda_{\mathrm{EP}}) = E_{\mathrm{EP}},
    \qquad
    \end{equation}
    \begin{equation}       
    |\psi_1(\lambda_{\mathrm{EP}})\rangle = |\psi_2(\lambda_{\mathrm{EP}})\rangle
    = |\psi_{\mathrm{EP}}\rangle,
\end{equation}
making the Hamiltonian non-diagonalizable. In a non-Hermitian setting the
eigenpairs satisfy
\begin{equation}
    H(\lambda)|\psi_n\rangle = E_n(\lambda)|\psi_n\rangle, \qquad
    \langle\tilde{\psi}_n|H(\lambda) = \langle\tilde{\psi}_n|E_n(\lambda),
\end{equation}
where $E_n$ is, in general, complex and away from an EP one imposes the
biorthogonal normalization $\langle\tilde{\psi}_i|\psi_j\rangle =
\delta_{ij}$~\cite{OKOLOWICZ2003271}. A convenient measure of the departure
from this normalization is the \emph{phase rigidity}
\begin{equation}
    r_i \equiv \frac{\langle\tilde{\psi}_i|\psi_i\rangle}
    {\langle\psi_i|\psi_i\rangle},
\end{equation}
which equals unity for well-separated resonances and vanishes at the EP,
$\langle\tilde{\psi}_{\mathrm{EP}}|\psi_{\mathrm{EP}}\rangle = 0$, as the
states become self-orthogonal through their mutual coupling to the continuum.
The vanishing of $r_i$ has been widely used as a practical indicator of an
EP~\cite{PhysRevX.6.021007, Jaiswal_2023}, with its power-law scaling $r_i \sim
\varepsilon^{(n-1)/n}$ \cite{PhysRevResearch.5.033042, Heiss_2008} characterizing the order $n$ of the singularity, though
deviating exponents of the form $(n-1)/2$ and $n-1$ have been encountered in anisotropic EPs \cite{PhysRevB.99.241403}.



The self-orthogonality condition 
$\langle\tilde{\psi}_{\mathrm{EP}}|\psi_{\mathrm{EP}}\rangle = 0$ at the 
exceptional point~\cite{Kato1966,Heiss_2012} has a direct spectral consequence: 
it causes individual eigenstate expectation values to diverge. Nevertheless, we 
show that the sum over both modes remains finite, with the divergences 
cancelling exactly due to the structure of the Puiseux expansion.

Setting $\varepsilon = \lambda - \lambda_{\mathrm{EP}}$
the eigenvalues and eigenvectors admit the Puiseux 
expansions~\cite{Kato1966,Heiss2004}
\begin{align}
E_\pm &= E_{\mathrm{EP}} \pm c_E\,\varepsilon^{1/2} + O(\varepsilon), \\
|\psi^{(\pm)}\rangle &= |\psi_{\rm{EP}}\rangle \pm \varepsilon^{1/2}|\psi_1\rangle 
   + O(\varepsilon), \\
\langle\tilde{\psi}^{(\pm)}| &= \langle\tilde{\psi}_{\rm{EP}}| 
   \pm \varepsilon^{1/2}\langle\tilde{\psi}_1| + O(\varepsilon),
\end{align}
where $|\psi_1\rangle$ and $\langle\tilde{\psi}_1|$ are the right and left 
generalised eigenvectors forming the Jordan chain at the 
EP~\cite{Heiss_2012,Moiseyev_2011}. The $\pm$ signs labelling the two branches 
originate from the square-root topology of the eigenvalue surface near the 
exceptional point~\cite{Heiss1999,Dembowski2001}.

The expectation value of an observable $\hat{O}$ in eigenstate 
$|\psi^{(\pm)}\rangle$ is given by the biorthogonal 
expression~\cite{Brody2014}
\begin{equation}
\langle\hat{O}\rangle^{(\pm)} 
= \frac{\langle\tilde{\psi}^{(\pm)}|\hat{O}|\psi^{(\pm)}\rangle}
       {\langle\tilde{\psi}^{(\pm)}|\psi^{(\pm)}\rangle}.
\end{equation}
We now expand the numerator and denominator separately using the Puiseux series.

Inserting the expansions into the numerator and collecting terms up to order 
$\varepsilon^{1/2}$ gives
\begin{equation}
\langle\tilde{\psi}^{(\pm)}|\hat{O}|\psi^{(\pm)}\rangle 
= \gamma \pm \varepsilon^{1/2}\,\delta + O(\varepsilon),
\end{equation}
where we have defined
\begin{align}
\gamma &\equiv \langle\tilde{\psi}_{\rm{EP}}|\hat{O}|\psi_{\rm{EP}}\rangle, \\
\delta &\equiv \langle\tilde{\psi}_{\rm{EP}}|\hat{O}|\psi_1\rangle 
            + \langle\tilde{\psi}_1|\hat{O}|\psi_{\rm{EP}}\rangle.
\end{align}

For the denominator, the self-orthogonality condition 
$\langle\tilde{\psi}_{\rm{EP}}|\psi_{\rm{EP}}\rangle = 0$ kills the leading term, so
\begin{equation}
\langle\tilde{\psi}^{(\pm)}|\psi^{(\pm)}\rangle 
= \pm\, c\,\varepsilon^{1/2} + O(\varepsilon),
\end{equation}
with
\begin{equation}
c \equiv \langle\tilde{\psi}_{\rm{EP}}|\psi_1\rangle 
      + \langle\tilde{\psi}_1|\psi_{\rm{EP}}\rangle \neq 0.
\end{equation}
The overlap is thus dominated by the $\varepsilon^{1/2}$ corrections. Crucially, 
the two modes have self-overlaps with opposite signs, a direct consequence of 
the $\pm$ structure in the Puiseux expansion.

Combining numerator and denominator, the expectation value in each eigenstate 
becomes
\begin{equation}
\langle\hat{O}\rangle^{(\pm)} 
= \frac{\gamma \pm \varepsilon^{1/2}\delta}{\pm\, c\,\varepsilon^{1/2}} 
  + O(\varepsilon^{1/2})
= \pm\frac{\gamma}{c\,\varepsilon^{1/2}} + \frac{\delta}{c} 
  + O(\varepsilon^{1/2}).
\end{equation}
Each eigenstate expectation value diverges individually as $\varepsilon \to 0$:
\begin{equation}
\langle\hat{O}\rangle^{(\pm)} \xrightarrow{\;\varepsilon\to 0\;} \pm\infty.
\end{equation}
However, the opposite signs in the divergent contributions ensure exact cancellation when summing over both modes:
\begin{equation}
\langle\hat{O}\rangle^{(+)} + \langle\hat{O}\rangle^{(-)} 
= \frac{2\delta}{c} + O(\varepsilon^{1/2}),
\end{equation}
where the $\pm\gamma/(c\varepsilon^{1/2})$ terms cancel exactly. Taking the 
limit $\varepsilon \to 0$, we obtain the finite result
\begin{multline}
\langle\hat{O}\rangle^{(+)} + \langle\hat{O}\rangle^{(-)} 
\xrightarrow{\;\varepsilon\to 0\;} \\
\frac{2}{c}\Bigl(
\langle\tilde{\psi}_{\rm{EP}}|\hat{O}|\psi_1\rangle
+ \langle\tilde{\psi}_1|\hat{O}|\psi_{\rm{EP}}\rangle
\Bigr).
\end{multline}
The finite remainder is expressed entirely through the coalesced eigenvector 
$|\psi_{\rm{EP}}\rangle$ and the generalized eigenvector $|\psi_1\rangle$, confirming 
that the Jordan structure of $H$ at the exceptional point fully determines the 
physical observables~\cite{Heiss_2012,Moiseyev_2011}.

An EP can also be viewed as a double pole of the S-matrix. Let's consider one resonance of energy $E_r$ and width $\Gamma$ coupled to two decay channels $c=1,2$ of partial widths  $\Gamma_1$ and $\Gamma_2$. The S-matrix for the elastic scattering for the channel $c=1$ can be written as 
\begin{equation}
    S_{11}(E)
=
\frac{E - E_r - \tfrac{i}{2}\left(\Gamma_1 - \Gamma_2\right)}
     {E - E_r + \tfrac{i}{2}\left(\Gamma_1 + \Gamma_2\right)}
\end{equation}
the corresponding elastic cross section gives

\begin{align}
    \sigma_{11}(E)
&= \frac{\pi}{k_1^2} |1-S_{11}(E)|^2\\ \label{ecs}
    &=
\frac{\pi}{k_1^2}  \frac{\Gamma_1^2}
     {(E - E_r)^2 + \left(\Gamma/2\right)^2}
\end{align}
where $\Gamma=\Gamma_1+\Gamma_2$. At the EP, the S-matrix for the elastic scattering of channel $c=1$ takes the form

\begin{equation}
    S_{11}(E)
=
\frac{(E - E_{EP} - \frac{i}{2}\left(\Gamma_1 - \Gamma_2\right))^2}
     {(E - E_{EP} + \frac{i}{2}\left(\Gamma_1 + \Gamma_2\right))^2},
\end{equation}
where $E_{EP}$ is the energy of the resonance at the EP. Inserting this expression into (\ref{ecs}), one obtains the elastic cross section for the channel $c=1$ at the EP as

\begin{equation}
    \sigma_{11}(E)= \frac{4\pi}{k_1^2} \frac{(E - E_{EP})^2 + \left(\Gamma_2/2\right)^2}
     {(E - E_{EP})^2 + \left(\Gamma/2\right)^2} \frac{\Gamma_1^2}
     {(E - E_{EP})^2 + \left(\Gamma/2\right)^2},\label{cs1}
\end{equation}
which resembles the result for a normal resonance multiplied by an extra interference factor coming from the double pole that splits the peak in half at $E=E_{EP}$. Similarly, for the channel $c=2$ one obtains
\begin{equation}
    \sigma_{22}(E)= \frac{4\pi}{k_2^2} \frac{(E - E_{EP})^2 + \left(\Gamma_1/2\right)^2}
     {(E - E_{EP})^2 + \left(\Gamma/2\right)^2} \frac{\Gamma_2^2}
     {(E - E_{EP})^2 + \left(\Gamma/2\right)^2}.\label{cs2}
\end{equation}
Contrary to the single channel case, where the cross section is zero at the minimum of the valley, as shown in \cite{cardona}, the ratio between the partial widths in the two channel case defines how deep this valley is or if there is one at all.

\section{Results} \label{sec:results}
In this section we apply the GSM-CC framework to three light nuclei in which an exceptional point is located by tuning the $\ell=1$ spin-orbit strengths of the one-body potential for protons and neutrons. The cases are ordered by increasing channel complexity at the EP. Section~\ref{sec:A6} treats the $2^+$ doublet of $^6$Li, where the EP lies below the proton threshold and only the deuteron channel is open, providing the cleanest single-channel phenomenology. Section~\ref{sec:A7} addresses the $5/2^-$ doublet of the mirror pair $^7$Li and $^7$Be, where in $^7$Be two channels are open at the EP and the ratio of partial widths controls the channel-dependent fate of the split-peak signature. Section~\ref{sec:A8} examines the $2^+$ isospin doublet of $^8$Be, where the EP emerges only above the proton threshold through a channel-induced restructuring of the eigenstates.

\subsection{\texorpdfstring{$^6$\rm Li}{6Li}}\label{sec:A6}

The nucleus in question was modeled by $^4\text{He}$ core with 2 valence nucleons and using a channel basis comprising $[^4\text{He}(0_1^+) \otimes  {}^2\text{H}(L_j)]^{J^\pi}$,  $[^5\text{He}(K_i^\pi) \otimes p(l_j)]^{J^\pi}$ and $[^5\text{Li}(K_i^\pi) \otimes n(l_j)]^{J^\pi}$ channels. The cluster channels were constructed by coupling the partial waves  $L_j $= ${^3S}_{1}$, ${^3P}_{0}$, ${^3P}_{1}$, ${^3P}_{2}$, ${^3D}_{1}$, ${^3D}_{2}$, ${^3D}_{3}$ of the wave function of $^2\text{H}$ with the inert core $^4\text{He}$ in the ground state $0^+$. The one-proton and one-neutron channels were built by coupling the partial waves $l_j = s_{1/2}$, $p_{1/2}$, $p_{3/2}$, $d_{3/2}$, $d_{5/2}$, $f_{5/2}$, $f_{7/2}$ of their wave functions with the states $K_i^\pi= 3/2_1^-$, $1/2_1^-$ of $^5\text{He}$ and $^5\text{Li}$ respectively.

\begin{table}[ht]
\centering
\caption{Woods--Saxon potential parameters for protons and neutrons. 
The diffuseness $d = 0.65$~fm and radius $R_0 = 2.0$~fm are common to all partial waves. The central depth $V_0$ and spin-orbit strength $V_{so}$ are in MeV.}
\label{tab:WS_params}
\begin{tabular}{l c c}
\hline\hline
 & $V_0$  & $V_{so}$  \\
\hline
Proton ($\ell = 0$)      & 52.989 & 0 \\
Proton ($\ell \geq 1$)   & 52.989 & 3.492 \\
Neutron ($\ell = 0$)     & 54.088 & 0 \\
Neutron ($\ell \geq 1$)  & 54.088 & 3.936 \\
\hline\hline
\end{tabular}
\end{table}

\begin{table}[ht]
\centering
\caption{Nucleon--nucleon interaction parameters $V_0$ (in MeV) for the central, spin-orbit, and tensor components. The channels are labeled by spin $S$, isospin $T$, and parity (even/odd).}
\label{tab:NN_params}
\begin{tabular}{l c c}
\hline\hline
Channel & $(S,T)$ & $V_0$ (MeV) \\
\hline
\multicolumn{3}{c}{\textit{Central}} \\
\hline
Odd triplet   & $(1,1)$ & $0.498$ \\
Even triplet  & $(1,0)$ & $-3.921$ \\
Odd singlet   & $(0,0)$ & $-100.763$ \\
Even singlet  & $(0,1)$ & $206.923$ \\
\hline
\multicolumn{3}{c}{\textit{Spin-orbit}} \\
\hline
Odd triplet   & $(1,1)$ & $2.758$ \\
Even triplet  & $(1,0)$ & $-43.227$ \\
\hline
\multicolumn{3}{c}{\textit{Tensor}} \\
\hline
Odd triplet   & $(1,1)$ & $99.983$ \\
Even triplet  & $(1,0)$ & $-12.254$ \\
\hline\hline
\end{tabular}
\end{table}

The Hamiltonian consists of a one-body part of the Woods-Saxon (WS) type plus a spin-orbit term and a Coulomb field mimicking the core, and a nucleon-nucleon interaction of the Furutani-Horiuchi-Tamagaki (FHT) type~\cite{10.1143/PTP.62.981}.The parameters of this Hamiltonian, shown in Table~\ref{tab:WS_params} for the one-body potential and Table~\ref{tab:NN_params} for the two-body interaction, have been adjusted according to the data given in the ENSDF database~\cite{ENSDF}.
{\begin{figure}
    \centering
    \includegraphics[width=0.95\linewidth]{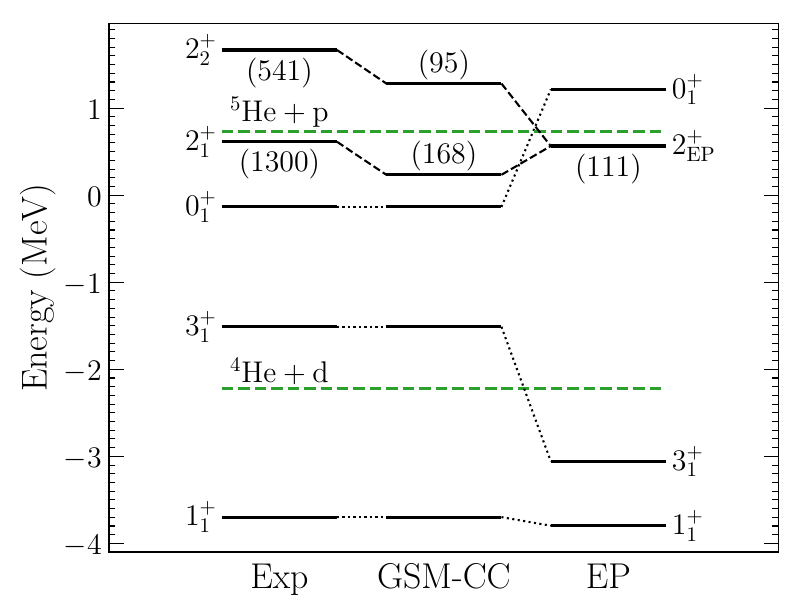}
    \caption{Experimental spectrum of $^6\text{Li}$ is compared with the GSM-CC spectrum.  The widths (in keV) of resonances are given in parentheses.  On the right-hand side, the spectrum for which the two $2^+$ states coalesce.}
    \label{spectrum_6Li}
\end{figure}
}
To account for the missing channels in Eq.~(\ref{react_channels}), the matrix elements for the channel-channel couplings involving one-nucleon reaction channels have been re-scaled by corrective factors $c(J^\pi):$ $c(1^+)= 1.0484$,  $c(3^+)=1.1$, $c(0^+)=1.0076$, $c(2^+)=0.9$. Similarly, for the matrix elements involving the deuteron channels, we have $c_{\text{d}}(3^+)=1.1$ and  $c_{\rm d}(2^+)=0.9$. 

\begin{figure}
    \centering
    \includegraphics[width=0.95\linewidth]{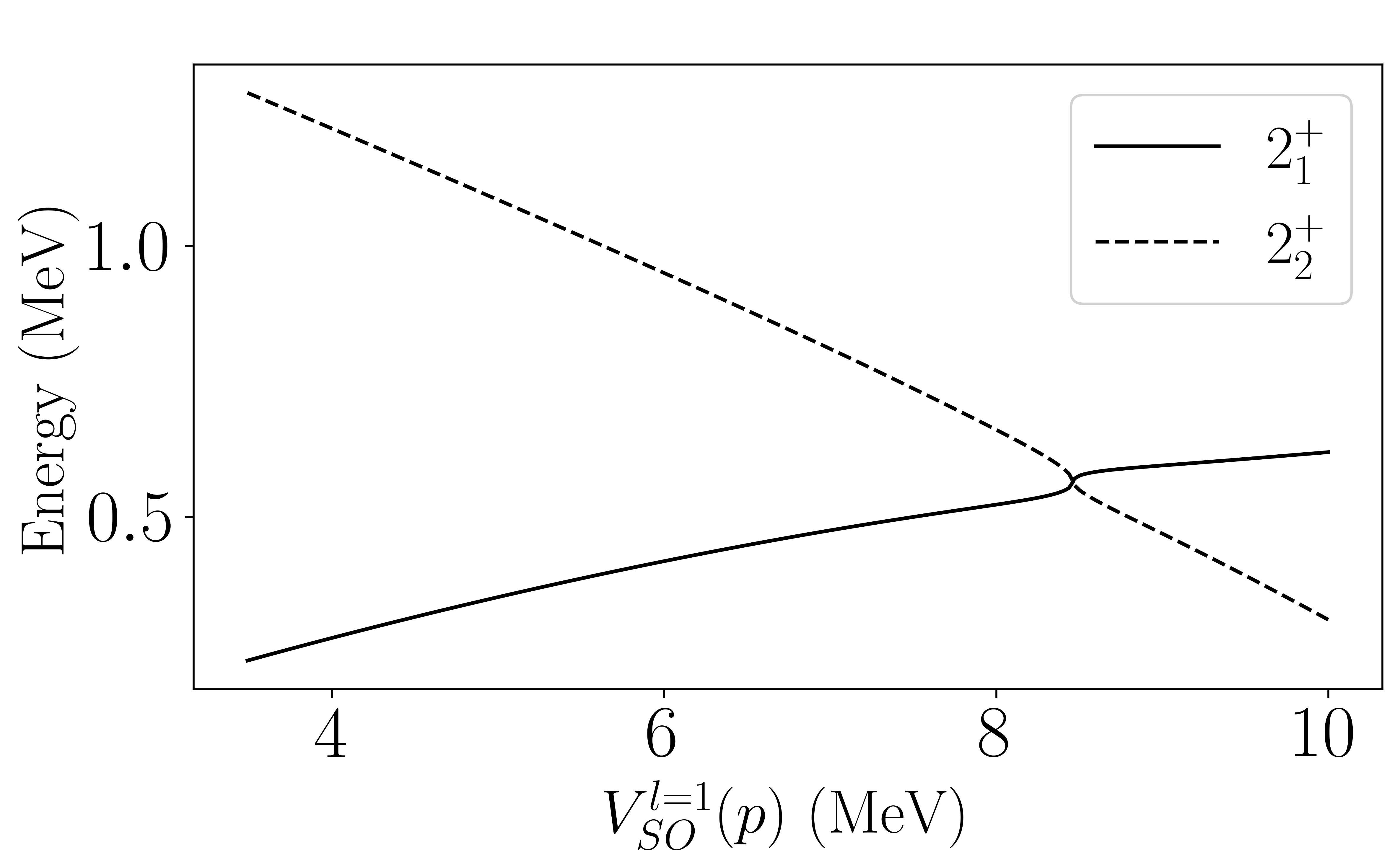}
        \includegraphics[width=0.95\linewidth]{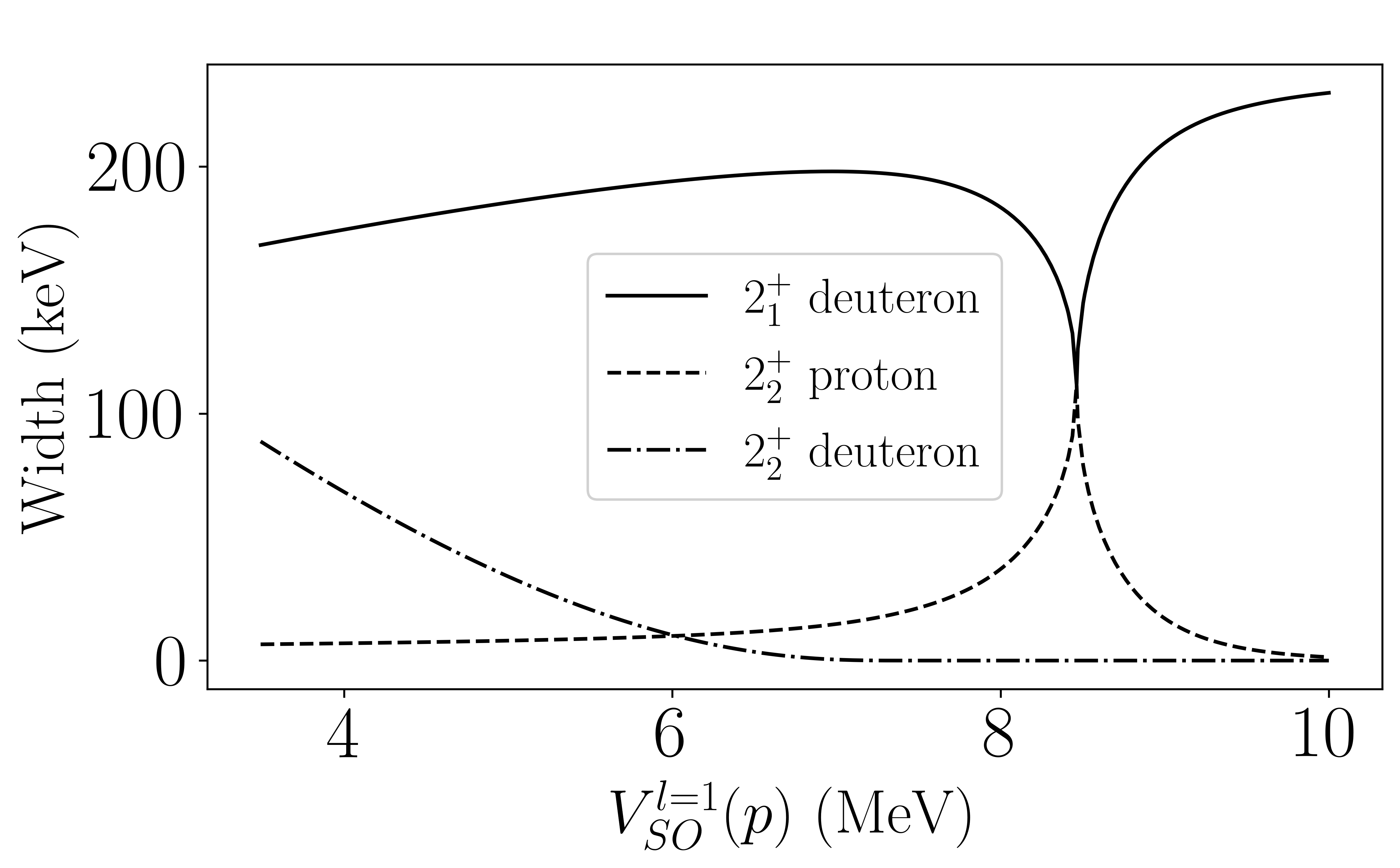}
    \caption{Trajectories of the $2^+$ doublet towards the EP, exhibiting the characteristic square-root behavior of energies and widths.}
    \label{E_6Li}
\end{figure}
Figure~\ref{spectrum_6Li} displays the experimental spectrum of $^6\text{Li}$ compared to the one obtained in GSM-CC and how the latter is modified at the EP, all relative to the binding energy of the $^4\text{He}$ core. Our calculation fairly reproduces the experimental spectrum. Although the $2^+$ are slightly lower in energy and their widths are underestimated, the separation between the states and the ratio between the widths are correctly reproduced. This allows for an accurate study of the EP present in this system, found using the strengths of $\ell=1$ spin-orbit one-body potential for protons and neutrons, $V^{(l=1)}_{SO}(p)$ and $V^{(l=1)}_{SO}(n)$,  as the variable parameters of the Hamiltonian $H(\lambda)$.

The evolution towards the EP of the energies and widths of the $2^+$ doublet is shown in Figure~\ref{E_6Li} as a function of the control parameters. With $2^+_1$ starting at $E_1=0.234$ MeV, $\Gamma_1 = 168$ keV and $2^+_2$ at $E_2=1.282$ MeV, $\Gamma_2=95$ keV, they coalesce at the EP found with energy $E_{\text{EP}}=0.564$ MeV and width $\Gamma_{\text{EP}}=111$ keV at $V^{(l=1)}_{SO}(p)=8.463$ MeV and $V^{(l=1)}_{SO}(n)=7.801$ MeV. Despite the fact that the coalescence of the $2^+$ doublet occurs between the two states, the other states are strongly modified. This shows that even if the states are close in energy, it doesn't mean that the EP is close in parameter space, as opposed to the $^7\text{Li}$ and $^7\text{Be}$ where the parameters had to be minimally modified to obtain the EP and their spectra remained almost unchanged.

\begin{figure}
    \centering
    \includegraphics[width=0.95\linewidth]{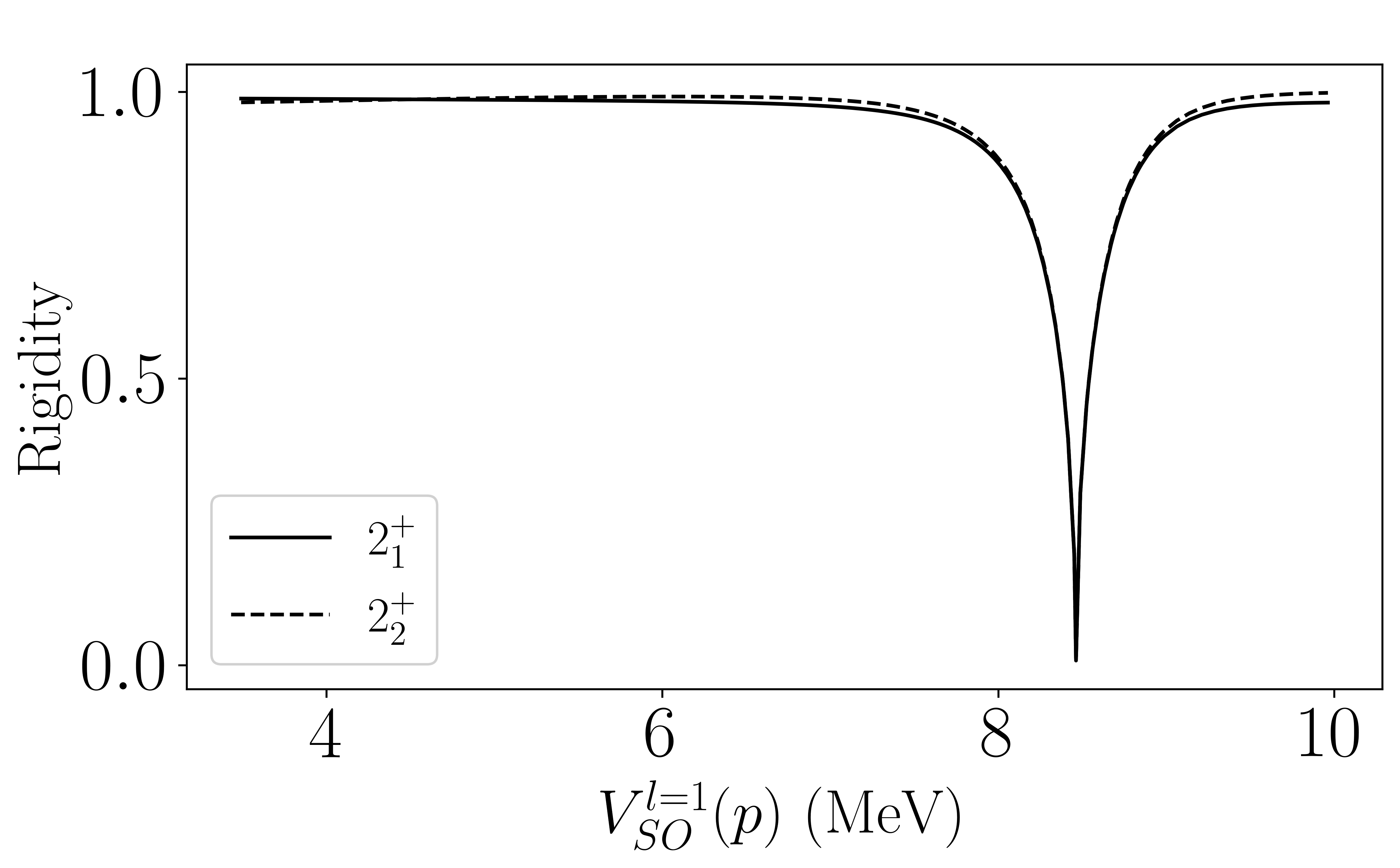}
    \caption{The phase rigidity of $2^+$ resonances is plotted as a function of the control parameters. The phase rigidity becomes equal to zero at the EP. }
    \label{6Li_rigidity}
\end{figure}

Figure~\ref{6Li_rigidity} illustrates the evolution of the phase rigidity for both $2^+$ states. Initially, both resonances have $r\approx1$, meaning that there is little to no mutual mixing, as these resonances do not overlap. As the states approach one another, their rigidity tends to zero, reflecting their growing non-orthogonality and the stronger mixing mediated by their shared decay channels. 

\begin{figure}
    \centering
    \includegraphics[width=0.95\linewidth]{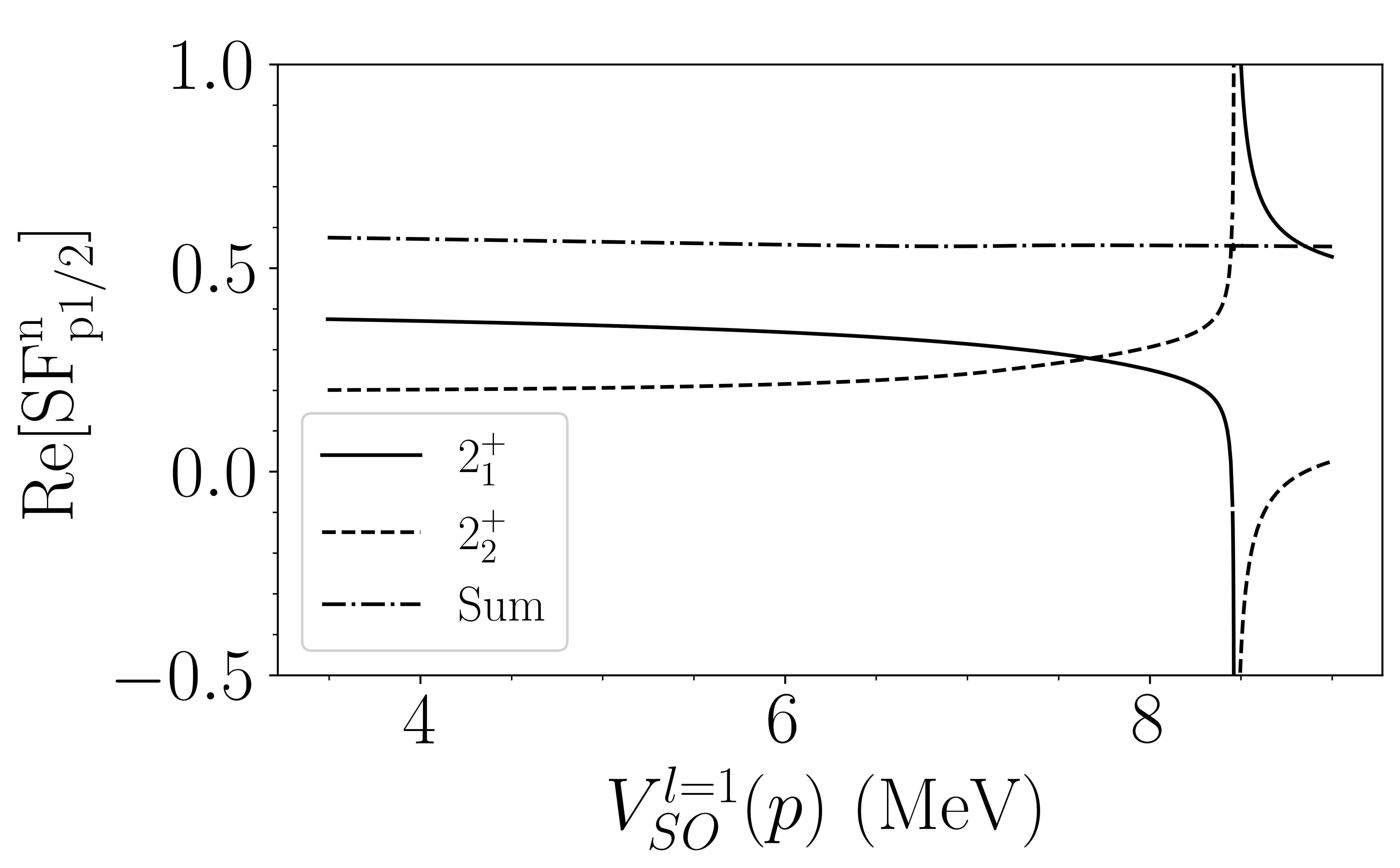}
    \includegraphics[width=0.95\linewidth]{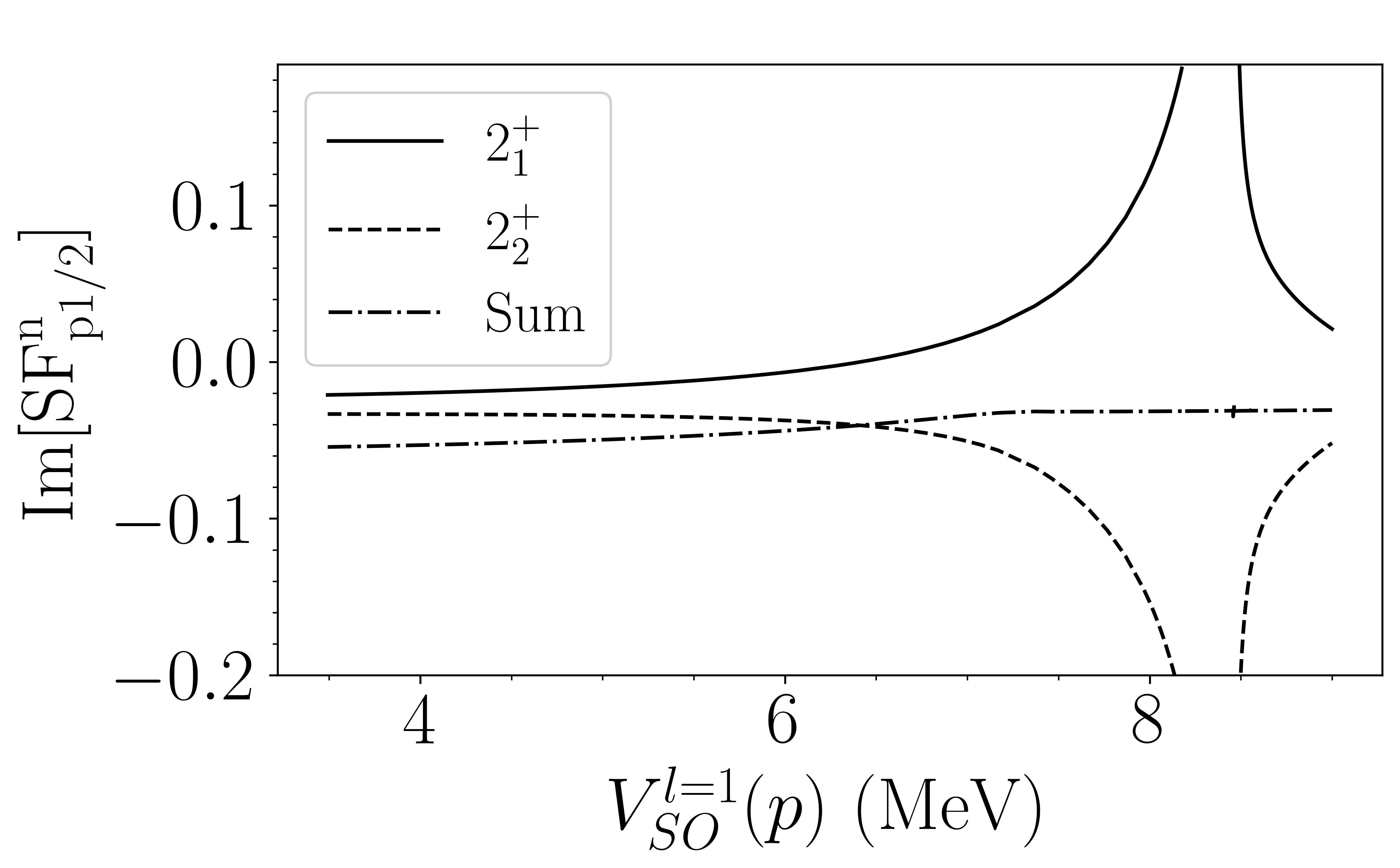}
    \caption{Evolution of the real part and imaginary part of the $\langle {}^6 \text{Li}(2^2_n) ||{}^5 \text{Li}(3/2^-)\otimes{} \text{n}(p_{1/2})\rangle$ spectroscopic factor for the $2^+_n$ ($n=1,2$) doublet of resonances in ${}^6 \text{Li}$ when approaching the EP as a function of the control parameters.}
    \label{SF_6Li}
\end{figure}

Related to the loss of orthogonality, Figure~\ref{SF_6Li} shows the evolution of spectroscopic factors for $\langle {}^6 \text{Li}(2^2_n) ||{}^5 \text{Li}(3/2^-)\otimes{} \text{n}(p_{1/2})\rangle$ as a function of the control parameters. At the beginning of the evolution, these quantities vary slowly, but as the states approach each other, the phase rigidity begins to vanish, and the states react strongly to even small parameter variations. At the EP, individual spectroscopic factors diverge as $S^{(\pm)} \sim 1/\sqrt{\epsilon}$, where $\epsilon$ measures the distance to coalescence. This divergence, arising from the vanishing biorthogonal norm, suggests an unphysical infinite coupling to decay channels and signals that individual eigenstates lose physical meaning at the EP. However, when the sum over both coalescing states is taken, these divergences compensate exactly: $S^{(+)} + S^{(-)}$ remains finite and continuous across the EP. This cancelation expresses the fact that, although individual eigenstates become ill-defined, the joined subspace they span remains physically meaningful. The total spectroscopic strength associated with the coalescing pair is well-behaved, varying smoothly as the Hamiltonian is modified by the control parameters.

\begin{figure}
    \centering
    \includegraphics[width=0.95\linewidth]{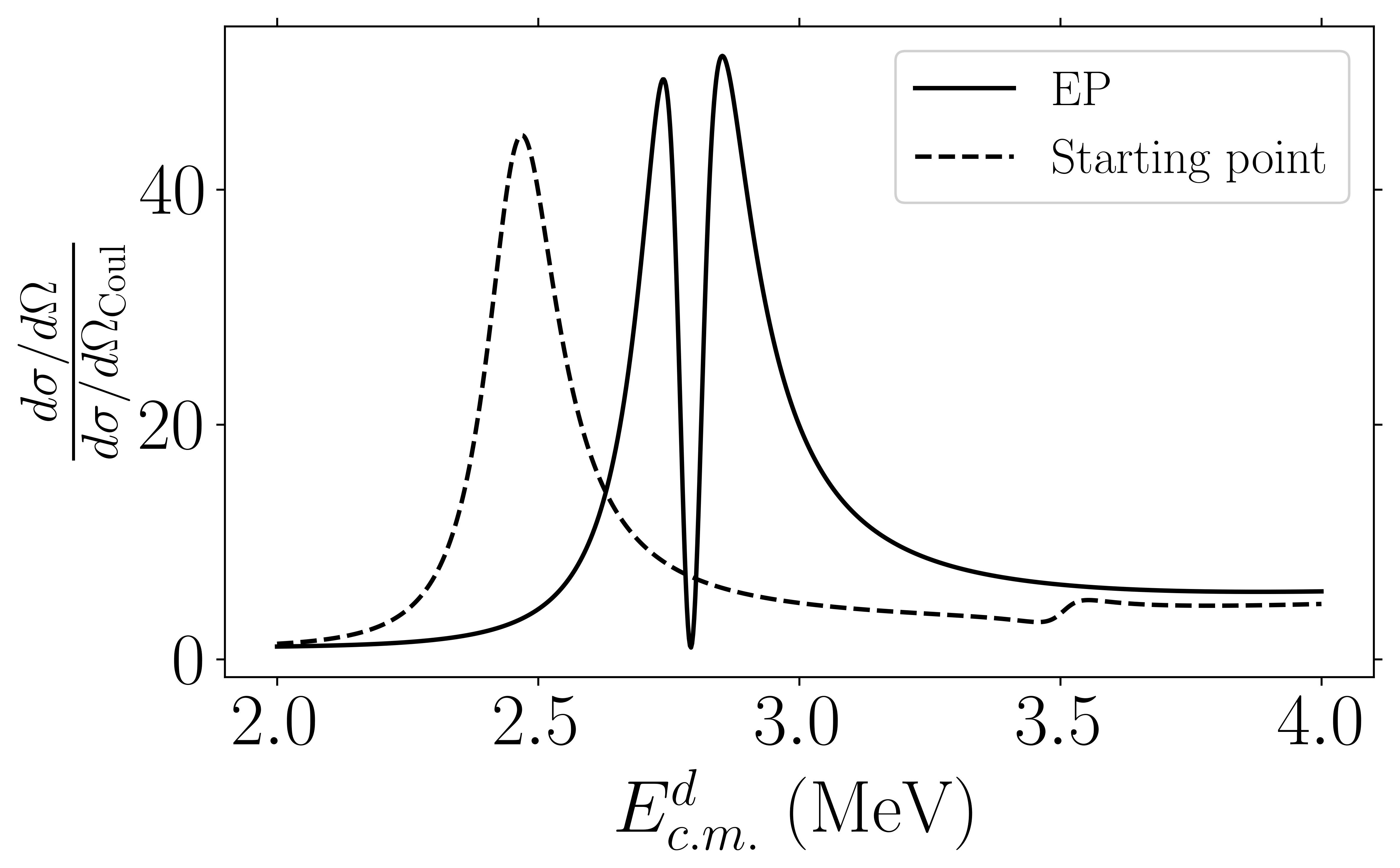}
    \caption{Elastic cross section for $^4\text{He}(d,d)^4\text{He}$ as function of the center of mass energy, showing the split peak behavior at the energy of the EP state. }
    \label{CS_6Li}
\end{figure}

Figure~\ref{CS_6Li} shows the contribution of the $2^+$ resonances to the elastic cross section for $^4\text{He}(d,d)^4\text{He}$. Since the EP state is located below the proton threshold, there is only one available channel, and its behavior follows the same structure as in \cite{cardona}, where the cross section shows a dramatic increase in the height of both peaks, which is, in reality, a split-peak with a minimum located at the energy of the double-pole of the S-matrix.

\begin{figure}
    \centering
    \begin{overpic}[width=0.95\linewidth]{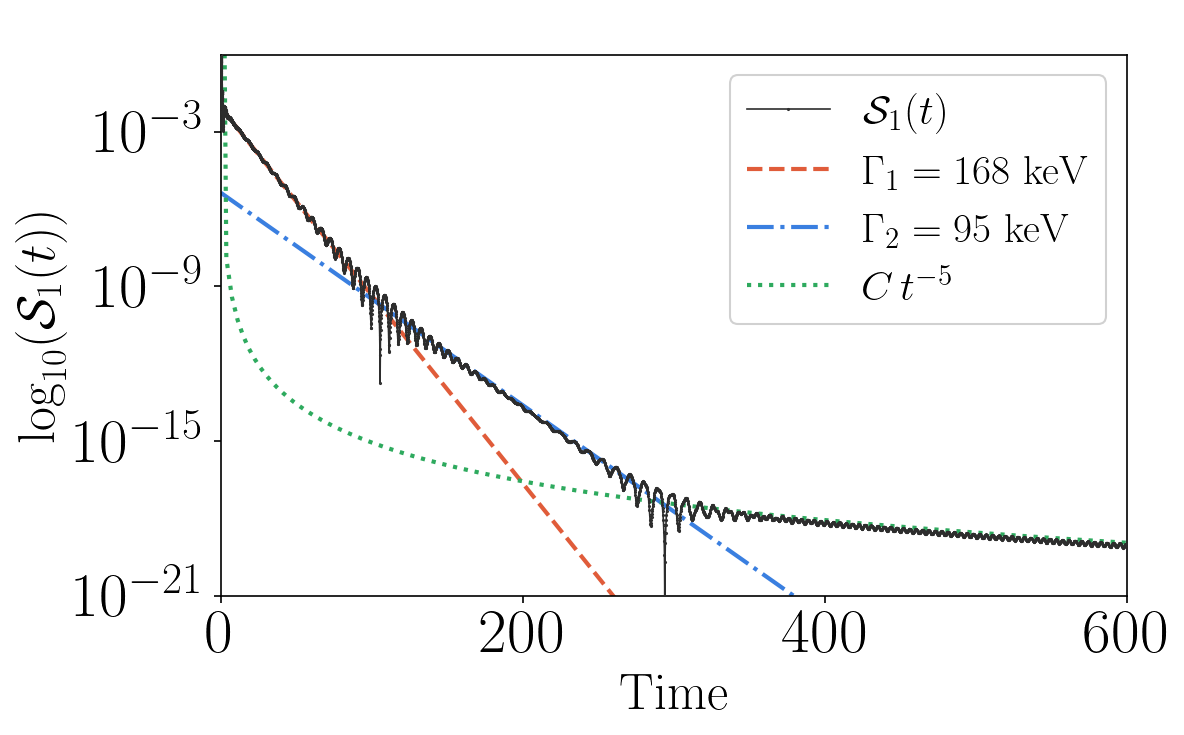}
        \put(22, 52){\large\textbf{a)}}
    \end{overpic}\\
    \begin{overpic}[width=0.95\linewidth]{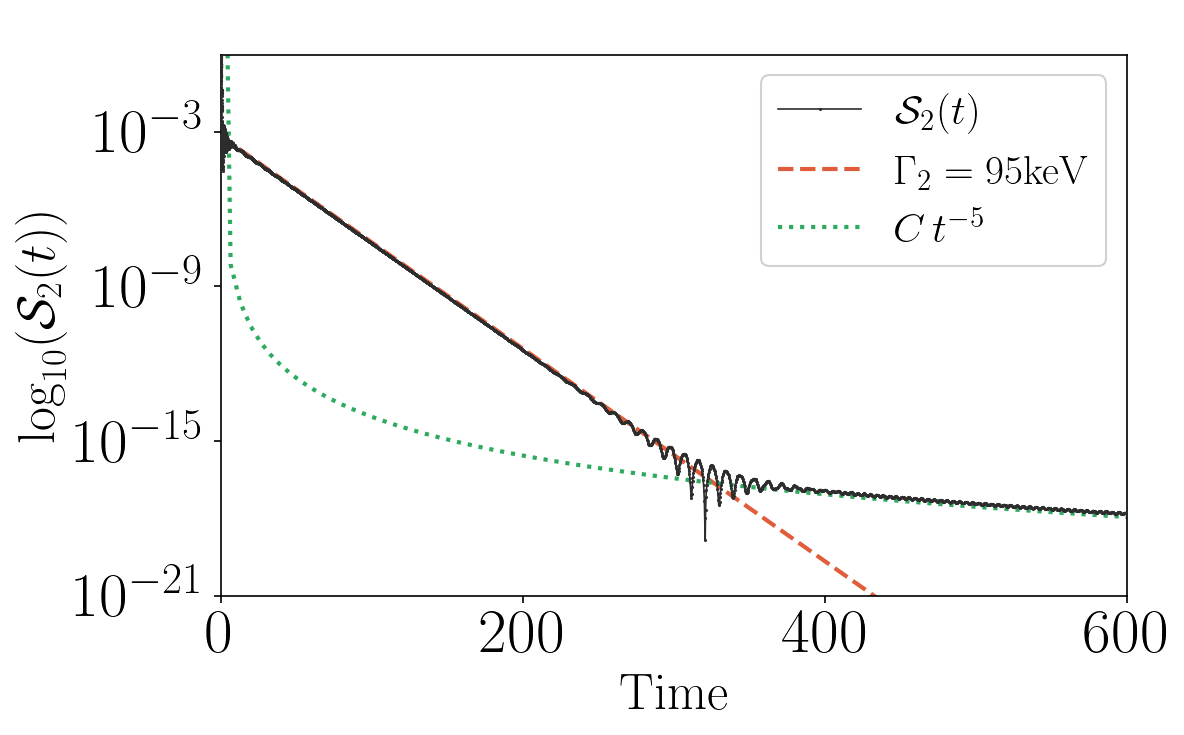}
        \put(22, 52){\large\textbf{b)}}
    \end{overpic}\\
    \begin{overpic}[width=0.95\linewidth]{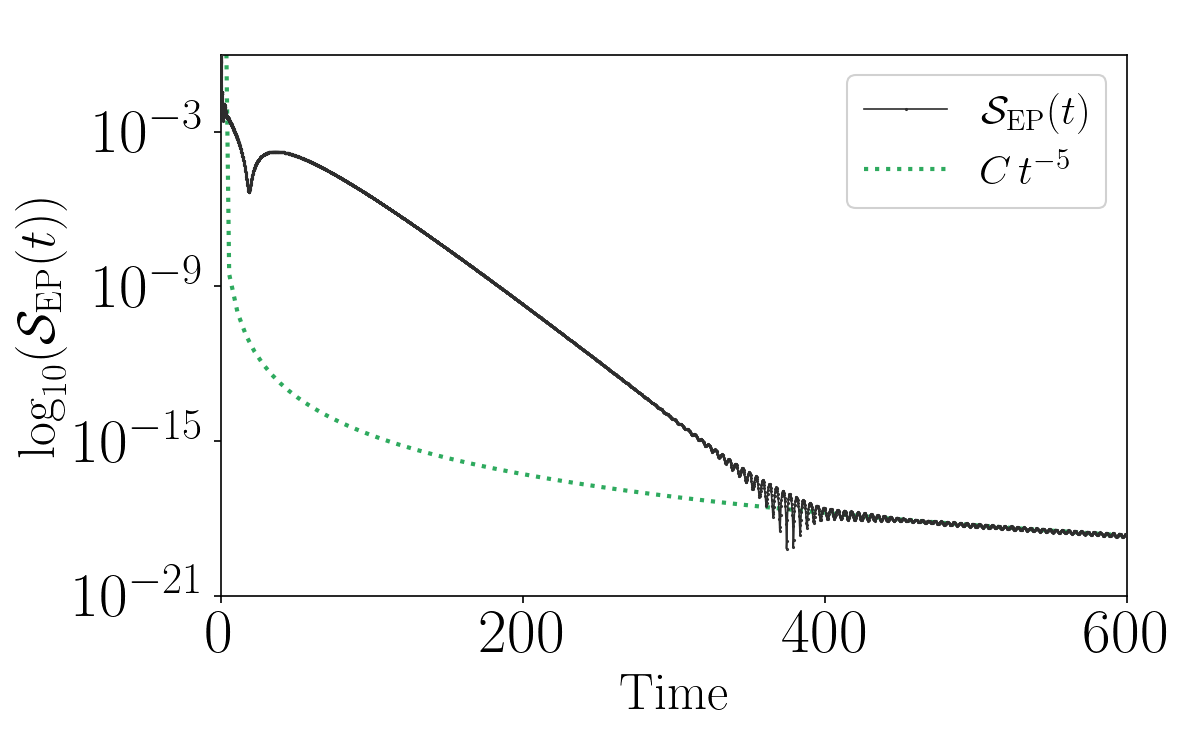}
        \put(22, 52){\large\textbf{c)}}
    \end{overpic}\\
    \begin{overpic}[width=0.95\linewidth]{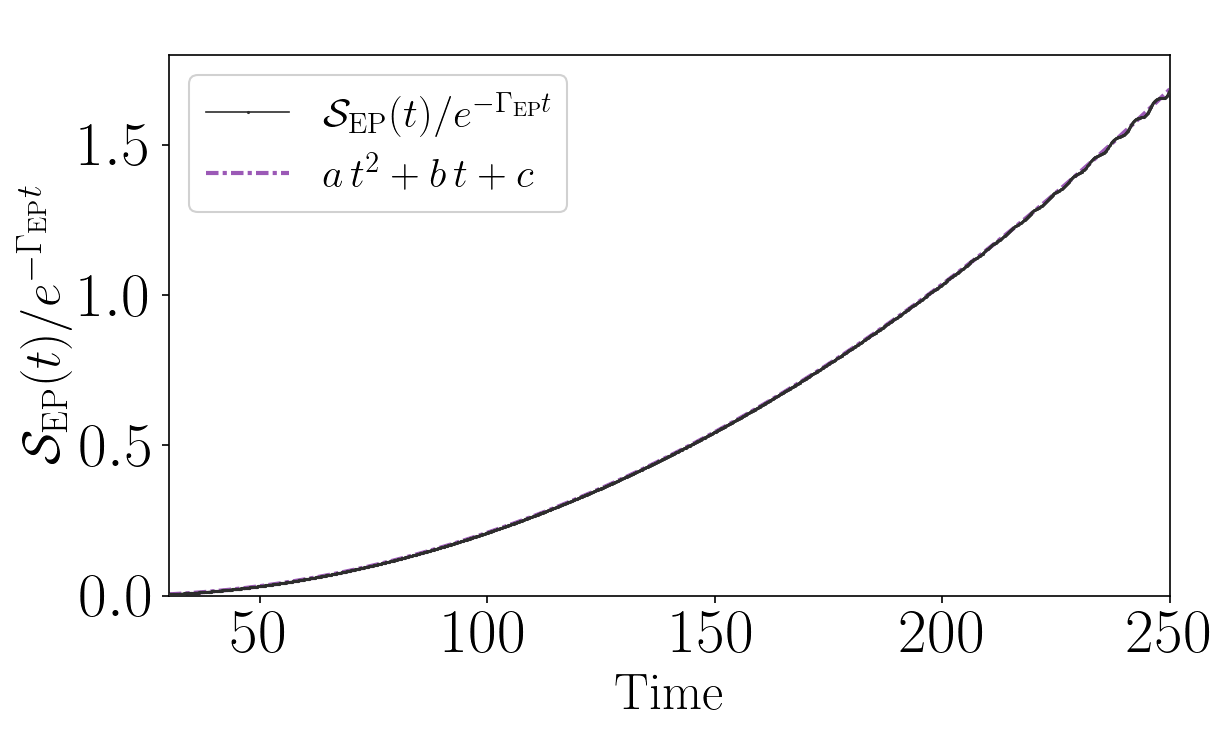}
        \put(85, 50){\large\textbf{d)}}
    \end{overpic}
    \caption{ Panels (a) and (b): Survival probabilities of the $2^+_1$ $2^+_2$  
    resonances at the starting point. Panel (c): Survival probability of the merged $2^+_{\rm EP}$ state at the exceptional point. Panel (d): Shows the intermediate region of the EP survival probability divided by $\exp(-\Gamma_{\rm EP} t)$ with $\Gamma_{\rm EP} = 111$ keV, together with a parabolic fit of the form $a + bt + ct^2$.}
    \label{sp_6Li}
\end{figure}

Figure~\ref{sp_6Li} shows the behavior of the survival probability $\mathcal{S}(t)$ of the studied $2^+$ states. The three expected regimes are clearly observed: a pre-exponential zone that depends on the form of the initial wave function, an intermediate region of exponential decay characterized by a decay coefficient equal to the widths of the states, and an inverse power-law part at long times proportional to $t^{-5}$.  This exponent comes from the relation $P(t) \approx t^{-(2l+3)}$ \cite{LFonda_1978, Peshkin_2014, PhysRevResearch.5.023183}, in this case the wave function is made up principally from $\ell=1$ poles. Each regime is separated by a transitional region of oscillatory behavior, where the contributions of both regimes are of the same order and interfere with each other.

The initial wave function at $t=0$ is constructed from the Gamow state of each
resonance, an eigenstate of the Hamiltonian at complex energy
$E = E_r - i\Gamma/2$ and complex momentum $k = k_r - i k_i$ with $k_i > 0$. Its
outgoing asymptotic form $\sim e^{ikr} = e^{ik_r r}\,e^{+k_i r}$ diverges
exponentially in $r$ and is not square-integrable in the usual sense, acquiring
a finite norm only through the Berggren regularization. Such a state cannot be
placed on a real-space grid and propagated with $e^{-iHt/\hbar}$, since its
divergent tail renders it ill-defined as an initial condition for real-time
evolution.

To obtain an admissible state, the pole wave function is multiplied by a Fermi
function,
\begin{equation}
    f(r) = \left[\,1 + \exp\!\left(\frac{r-R}{a}\right)\right]^{-1},
\end{equation}
which leaves the interior unchanged while smoothly suppressing the divergent
tail. The product $\psi_{\mathrm{pole}}(r)\,f(r)$ is then localized and
square-integrable, so that after renormalization it defines a legitimate wave
packet on the real axis. The radius $R$ and the diffuseness $a$ are both taken
large for complementary reasons: a large $R$ places the cutoff beyond the region
where the resonance carries its physical structure, while a large $a$ ensures a
gentle cutoff that keeps the real-momentum content concentrated near the pole.
Together these choices maximize the overlap
between the Gamow state and the initial wave function at $t=0$ and ensure that the initial
state is both normalizable and faithful to the original resonance.

These parameters are not arbitrary. Both the
spectral function and the survival probability are robust against changes of
$R$ and $a$, converging to a fixed line shape once these values are sufficiently
large. There exists, therefore, a neighbourhood of parameters within which the
results can be regarded as converged and independent of the numerical
preparation of the initial state, and the values used here are chosen within
this regime.

In panel (a) of Figure~\ref{sp_6Li}, where the resonances are well separated and do not overlap, the survival probability in the exponential regime reduces to an approximate superposition of independent decay modes, each governed by its own width. Although the states share a common coupling to the scattering continuum, their large energy separation suppresses any significant mixing or interference; thus, this coupling only enables independent decay without modifying their intrinsic widths or producing oscillatory beating effects. The presence of two exponential slopes in a single survival probability arises because the projection onto a given resonance is not strictly pure, it carries a small admixture of its partner, so the initial state is effectively a superposition of both resonances with a dominant amplitude on one and a much smaller amplitude on the other. The survival probability is then a sum of two exponentials weighted by these amplitudes squared, and which slope is visible at a given time depends on the competition between amplitude size and decay rate: the dominant own-state contribution governs the early exponential window, while at longer times it falls below the small but slower-decaying partner contribution, which then dictates the late-time slope. The transition between the two regimes is therefore not a beating effect but a simple crossover between two independent exponential terms, occurring at the time at which their contributions to the survival probability become equal. This crossover is only observable when the small-amplitude admixture corresponds to the longer-lived state; in the opposite case, panel (b), the subdominant component decays away faster than the dominant one and remains hidden beneath it, so a single exponential is seen throughout. As a result, an initial mixture evolves as a sum of exponentials, with the earlier-time dynamics dominated by the fastest-decaying component and longer-time behavior governed by the longest-lived state, corresponding to the pole closest to the real energy axis.

In panel (c) of Figure~\ref{sp_6Li} is the survival probability at the EP. Three distinct regimes still remain. Panel (d) of Figure~\ref{sp_6Li} shows the calculated survival probability in the intermediate region divided by an exponentially decaying function with the corresponding width of the EP state, $\Gamma_{EP} = 111$ keV. The resulting behavior fits perfectly with a parabola of the form $a+bt+ct^2$, meaning that the decay is not purely exponential, which coincides with the results obtained in \cite{10.1063/1.4983809, PhysRevA.84.013419, PhysRevE.75.027201}.

\begin{figure}
    \centering
    \begin{overpic}[width=0.95\linewidth]{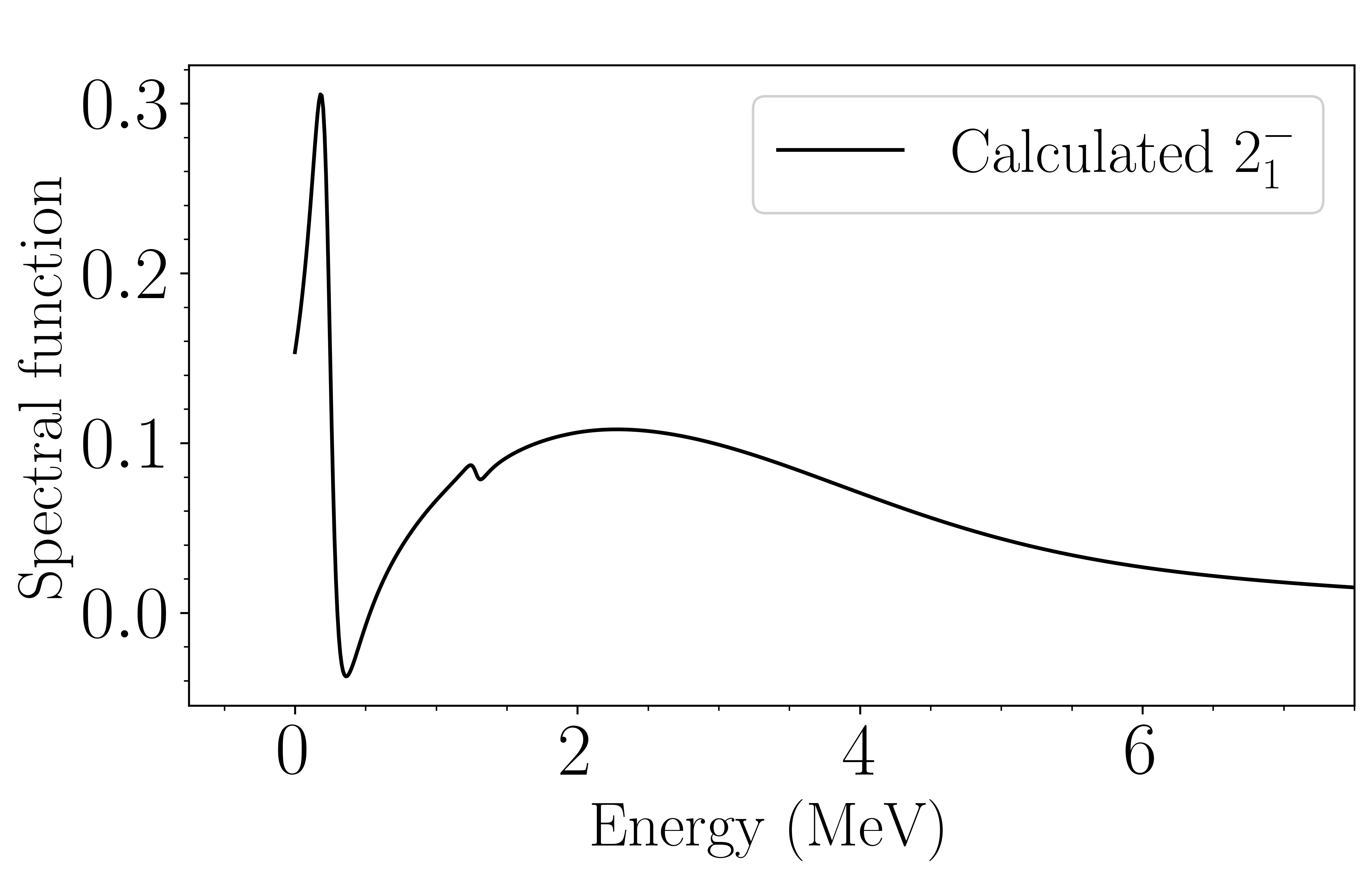}
        \put(16, 52){\large\textbf{a)}}
    \end{overpic}\\
    \begin{overpic}[width=0.95\linewidth]{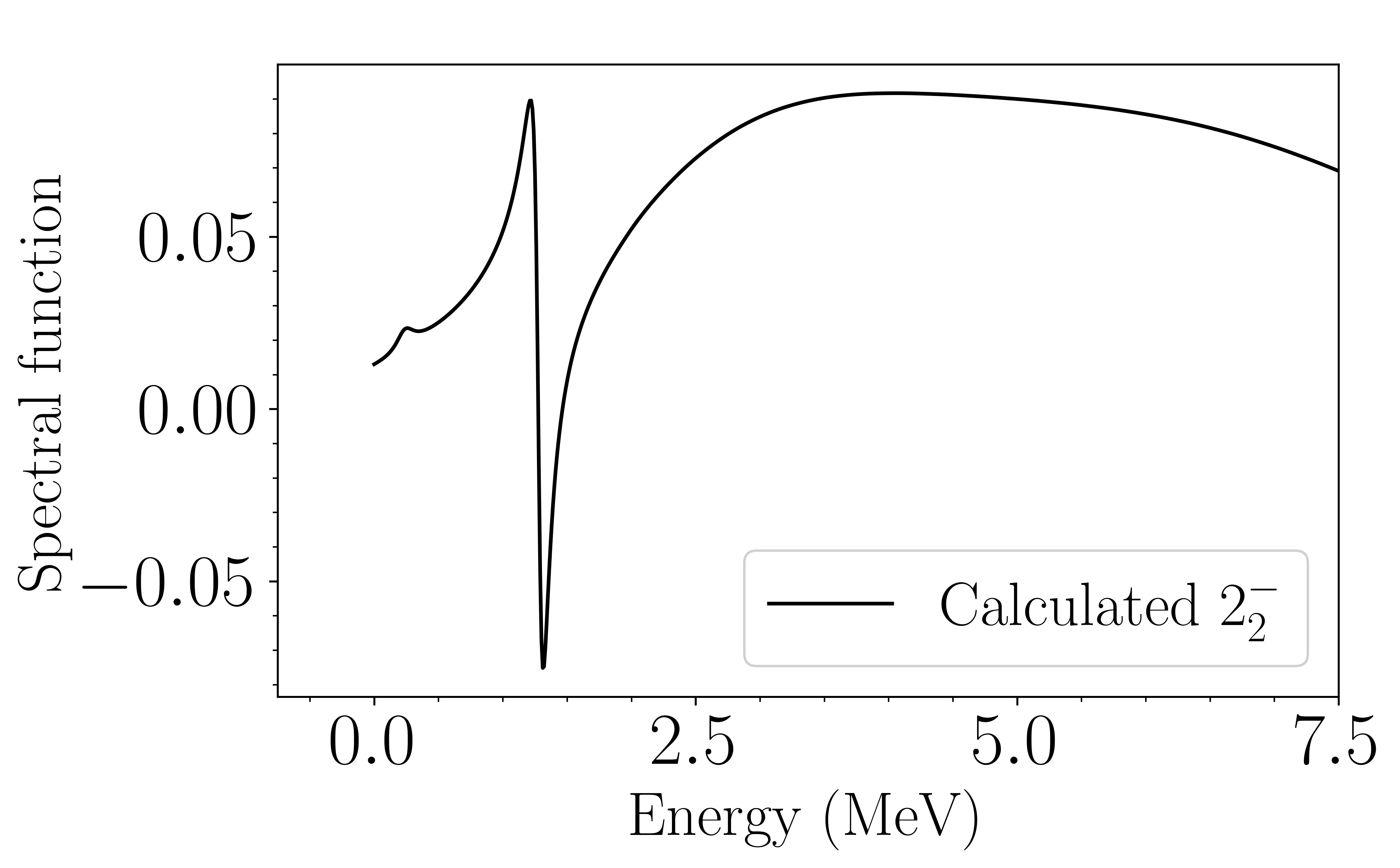}
        \put(22, 52){\large\textbf{b)}}
    \end{overpic}\\
    \begin{overpic}[width=0.95\linewidth]{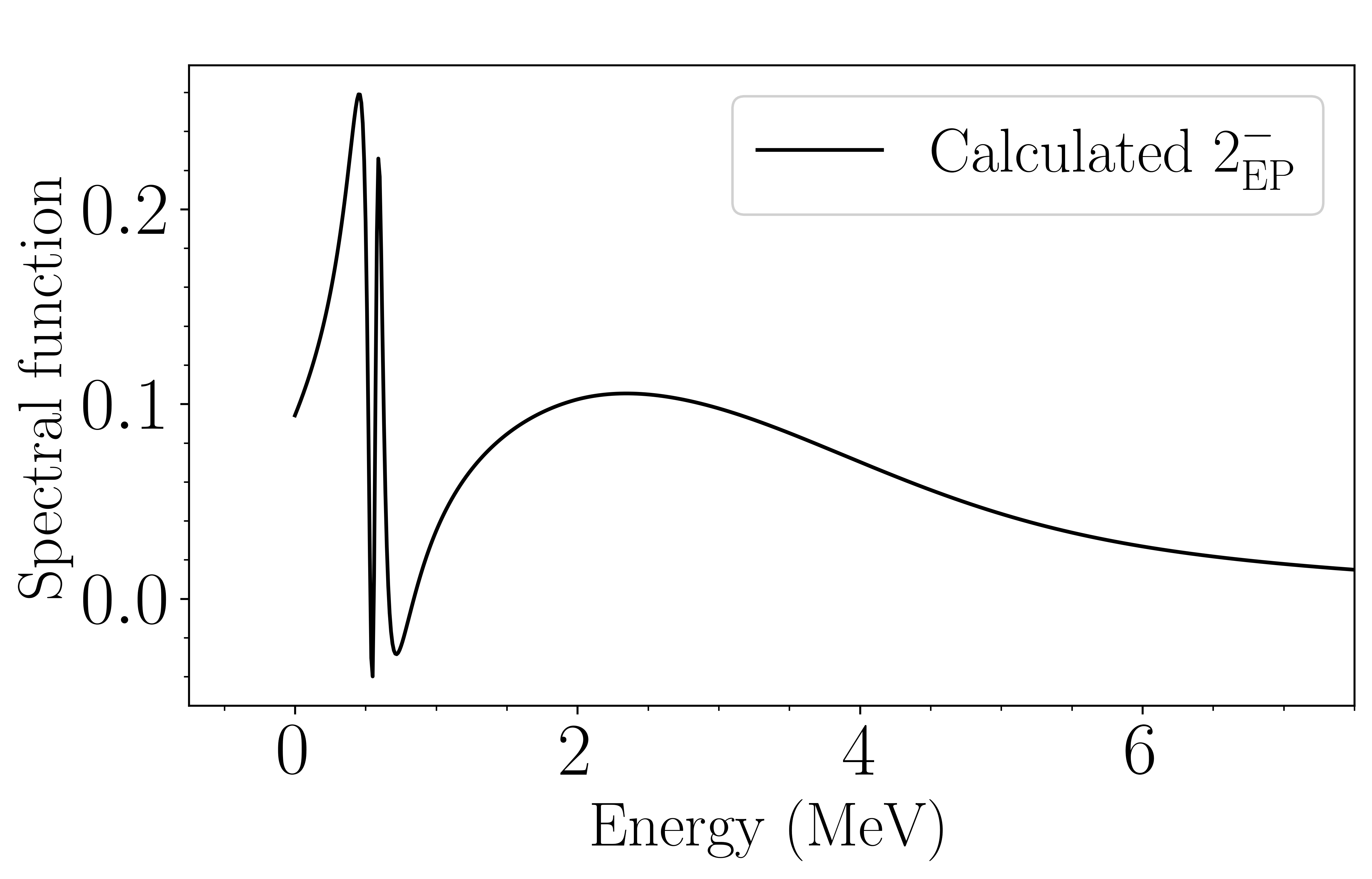}
        \put(16, 54){\large\textbf{c)}}
    \end{overpic}
    \caption{ Panels (a) and (b): Spectral functions of the $2^+_1$ and $2^+_2$ resonances at the starting point. Panel (c): Spectral function of the merged $2^+_{\rm EP}$ state at the exceptional point. Each panel shows a sharp resonance peak and a broad continuum bump at higher energies. At the EP, the two separate peaks merge into a single composite structure with two closely-spaced peaks around the EP energy.}
    \label{sfun_6Li}
\end{figure}

The corresponding spectral functions in Figure~\ref{sfun_6Li} support the same picture in the energy domain. The two sharp peaks sit at $0.234$ MeV and $1.282$ MeV, matching the calculated resonance energies $E(2^+_1) = 0.234$ MeV and $E(2^+_2) = 1.282$ MeV, with widths $\Gamma(2^+_1) = 168$ keV and $\Gamma(2^+_2) = 95$ keV. These widths fix the slopes of the exponential regimes of the survival probabilities, with the broader state decaying about $1.8$ times faster than the narrower one. Since the projection onto each resonance carries a small admixture of its partner, the small secondary peak visible in each spectral function is the energy-domain image of the subdominant exponential in $P(t)$. The inverted width hierarchy, the lower-energy state being the broader one, is precisely what makes the two-exponential crossover visible in the $2^+_1$ projection, where the partner admixture decays more slowly than the dominant component, and invisible in the $2^+_2$ projection, where it decays faster and stays hidden beneath the dominant exponential.Both spectral functions also display a broad bump extending several MeV above the resonance peaks, which reflects the non-resonant continuum components present in the initial state, not any additional resonance structure. These components dephase rapidly with respect to one another and produce the steep initial collapse of $P(t)$ before the resonance poles take over the dynamics; they therefore lie entirely outside the exponential regime that governs the long-time decay.

 \begin{figure}
    \centering
    \includegraphics[width=0.95\linewidth]{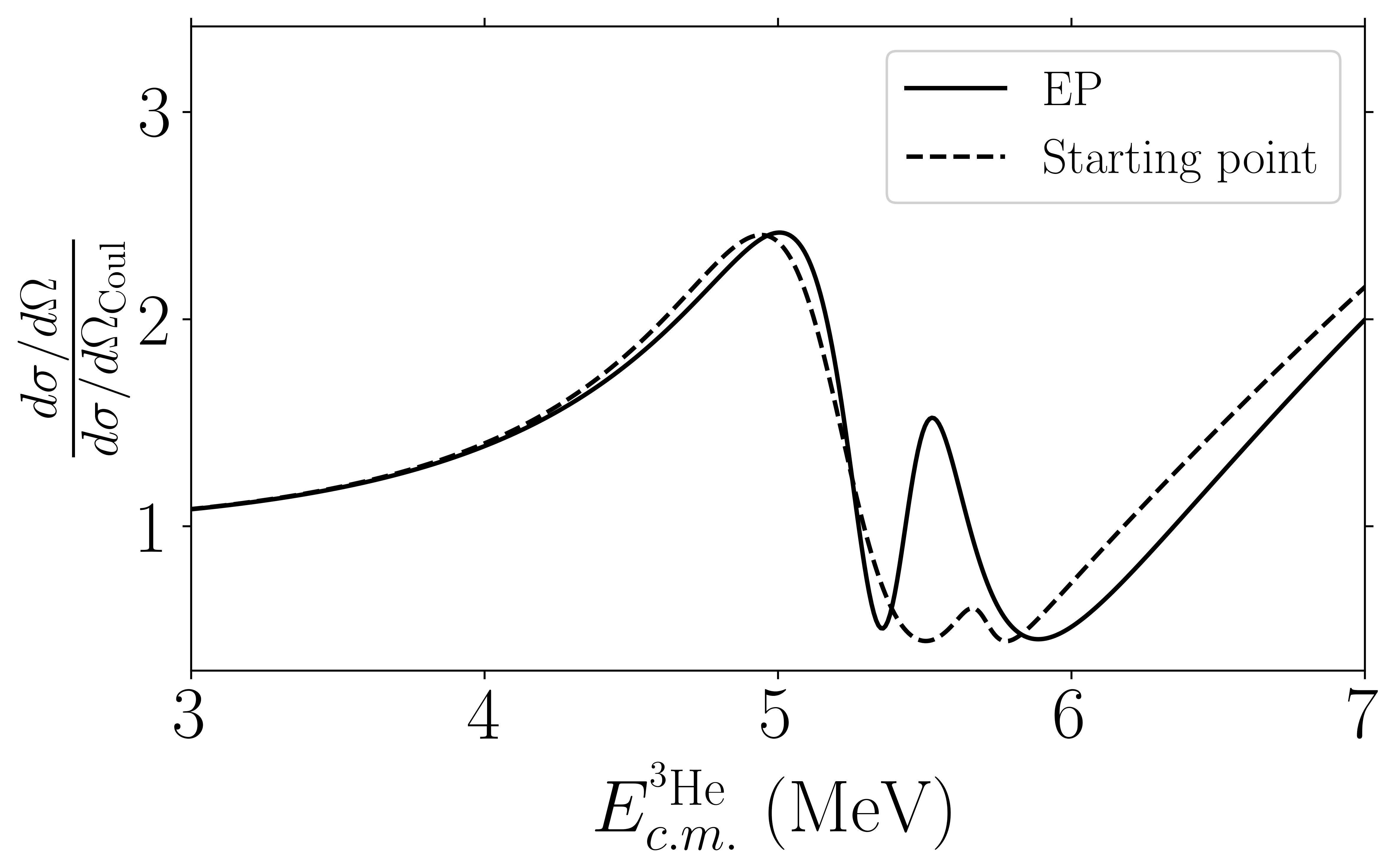}
        \includegraphics[width=0.95\linewidth]{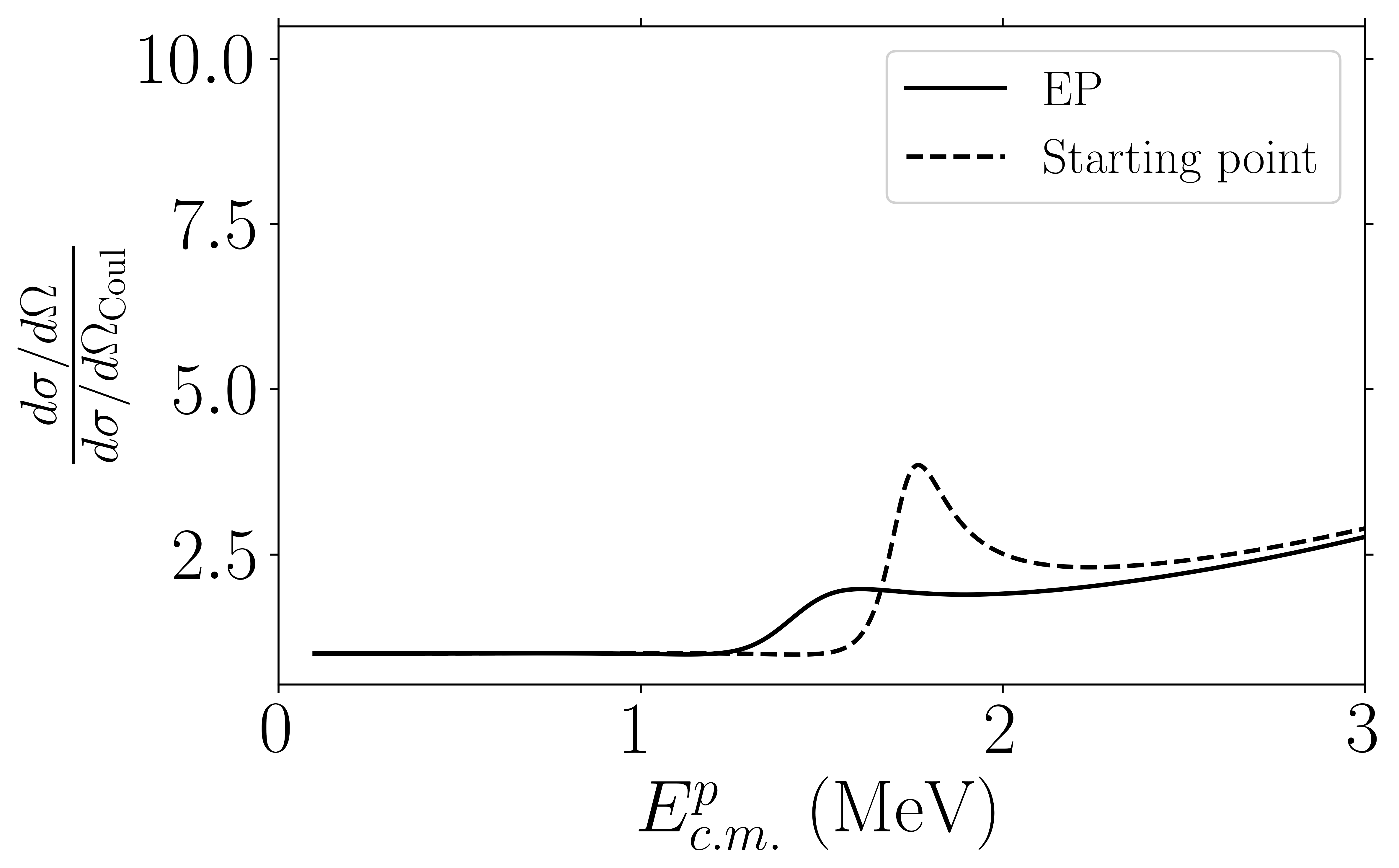}
    \caption{Elastic cross section for $^4\text{He}(^3\rm{He},^3\rm{He})^4\text{He}$ and $^6\text{Li}(p,p)^6\text{Li}$ as function of the center of mass energy, showing the effects of the EP when there are two channels available. The $^3\rm{He}$ channel is enhanced and shows a double peak structure, while the $p$ channel is suppressed.}
    \label{7Be_CS}
\end{figure}

At the EP, this same picture changes in a way that mirrors the time 
domain. The two well-separated peaks of the non-degenerate case have 
merged into a single composite structure with two closely-spaced lobes, the valley between them sitting at $0.565$ MeV, exactly the EP energy. 
The partner-admixture interpretation no longer applies: the small 
secondary peak that signaled the subdominant exponential at the starting 
point has grown to comparable height with the dominant one and merged 
with it, in direct correspondence with the disappearance of the two-
exponential crossover in $P(t)$. The non-resonant continuum bump after 
the merged feature is largely unchanged, confirming that the EP affects only the 
discrete resonance structure and not the short-time dynamics of $P(t)$.

\subsection{$^7{\rm Li}$ and $^{7}{\rm Be}$} \label{sec:A7}

In this section, we will discuss the effects of the EP associated with the $5/2^-$ doublet present in the spectra of the mirror nuclei $^7\text{Li}$ and $^7\text{Be}$. They were modeled by a $^4\text{He}$ core with 3 valence nucleons, and using a channel basis composed of $[^4\text{He}(0_1^+) \otimes {}^3\text{H}(L_j)]^{J^\pi}$, $[^5\text{He}(K_i^\pi) \otimes {}^2\text{H}(L_j)]^{J^\pi}$ and $[^6\text{Li}(K_i^\pi) \otimes n(l_j)]^{J^\pi}$ channels for $^7\text{Li}$, and $[^4\text{He}(0_1^+) \otimes {}^3\text{He}(L_j)]^{J^\pi}$, $[^5\text{Li}(K_i^\pi) \otimes {}^2\text{H}(L_j)]^{J^\pi}$ and $[^6\text{Li}(K_i^\pi) \otimes p(l_j)]^{J^\pi}$ channels for $^7\text{Be}$. 

The cluster channels related to the projectiles ${}^3\text{H}$ and ${}^3\text{He}$ were constructed by coupling the partial waves $L_j = S_{1/2}$, $P_{1/2}$, $P_{3/2}$, $D_{3/2}$, $D_{5/2}$, $F_{5/2}$, $F_{7/2}$ of their wave functions to the inert core $^4\text{He}$ in the ground state $0^+$. While the  channels related to $^2\text{H}$ were constructed by coupling the partial waves $L_j $= ${^3S}_{1}$, ${^3P}_{0}$, ${^3P}_{1}$, ${^3P}_{2}$, ${^3D}_{1}$, ${^3D}_{2}$, ${^3D}_{3}$ with the $^5\text{He}$ and $^5\text{Li}$ targets in the ground state $3/2^-$. Finally, one-nucleon channels were built by coupling the partial waves $l_j = s_{1/2}$, $p_{1/2}$, $p_{3/2}$, $d_{3/2}$, $d_{5/2}$, $f_{5/2}$, $f_{7/2}$ of the nucleon wave functions with the states $K_i^\pi= 1_1^+$, $3_1^+$, $0_1^+$ $2_1^+$, $2_2^+$, $1_2^+$ of $^6\text{Li}$.

The Hamiltonian parameters used for $^7\text{Li}$ and $^7\text{Be}$ are provided in \cite{cardona} and \cite{CardonaOchoa2026} respectively. Their spectra are fairly reproduced, as well as their widths. Hitherto, we have described the cases in which there is only one decay channel available for the EP state. In contrast to $^7\text{Li}$, in $^7\text{Be}$ the EP state is found above both the proton and ${}^3\text{He}$ thresholds. Figure~\ref{7Be_CS} shows both the elastic $^6$Li$(p,p)^6$Li and ${}^4\text{He}$(${}^3\text{He}$,${}^3\text{He}$)${}^4\text{He}$ cross sections at the starting point and at the EP. 

Although the double-pole structure of the S-matrix is still present in the 2-channel case, the cross section does not exhibit the split-peak behavior in all channels. The fact that the interference term in the cross sections depends on the partial decay width of the supplementary channel, as shown in Eqs.~(\ref{cs1}) and (\ref{cs2}), means that the ratio between the partial widths determines its effect on each cross section. Since the ${}^3\text{He}$ partial width is ten times the size of the proton's (see \cite{cardona}) in $5/2^-_{\text{EP}}$, the cross section for ${}^4\text{He}$(${}^3\text{He}$,${}^3\text{He}$)${}^4\text{He}$ is slightly enhanced and retains the split peak behavior, but the effect is not as strong as in the previous cases. On the contrary, the $^6$Li$(p,p)^6$Li cross section is suppressed compared to the starting point.   

\begin{figure}
    \centering      
    \begin{overpic}[width=0.95\linewidth]{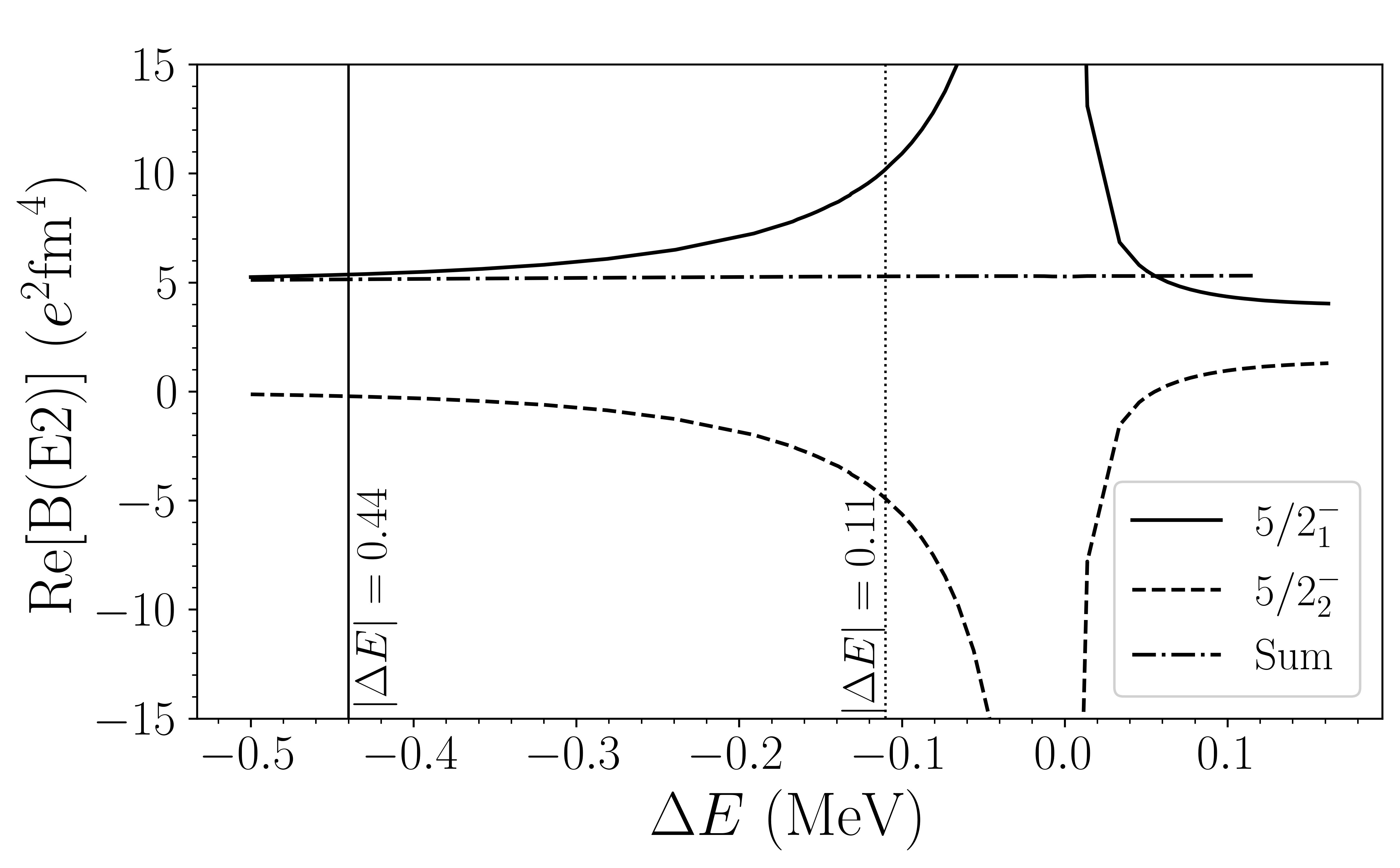}
        \put(20, 52){\large\textbf{a)}}
    \end{overpic}\\
    \begin{overpic}[width=0.95\linewidth]{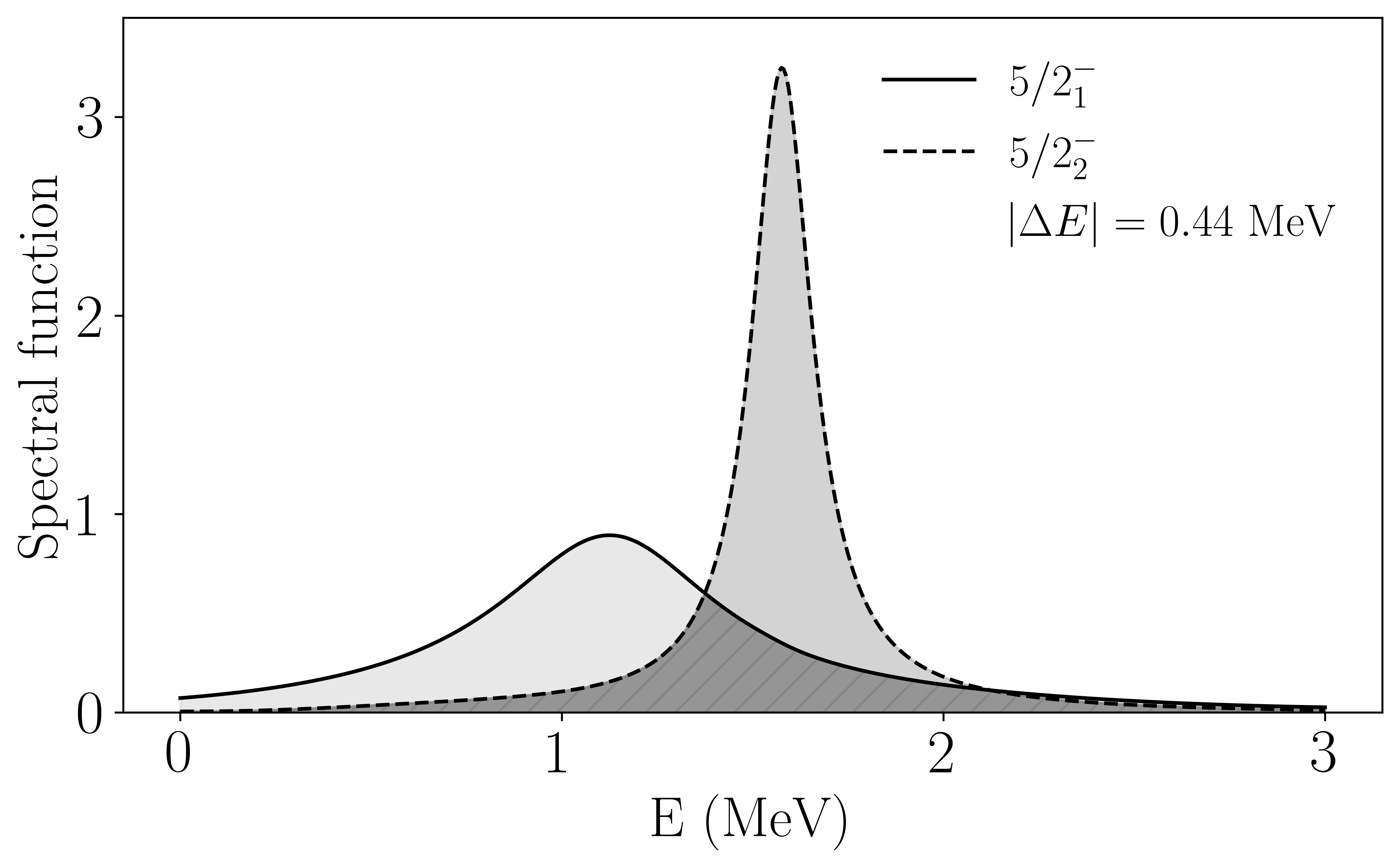}
        \put(12, 55){\large\textbf{b)}}
    \end{overpic}\\
    \begin{overpic}[width=0.95\linewidth]{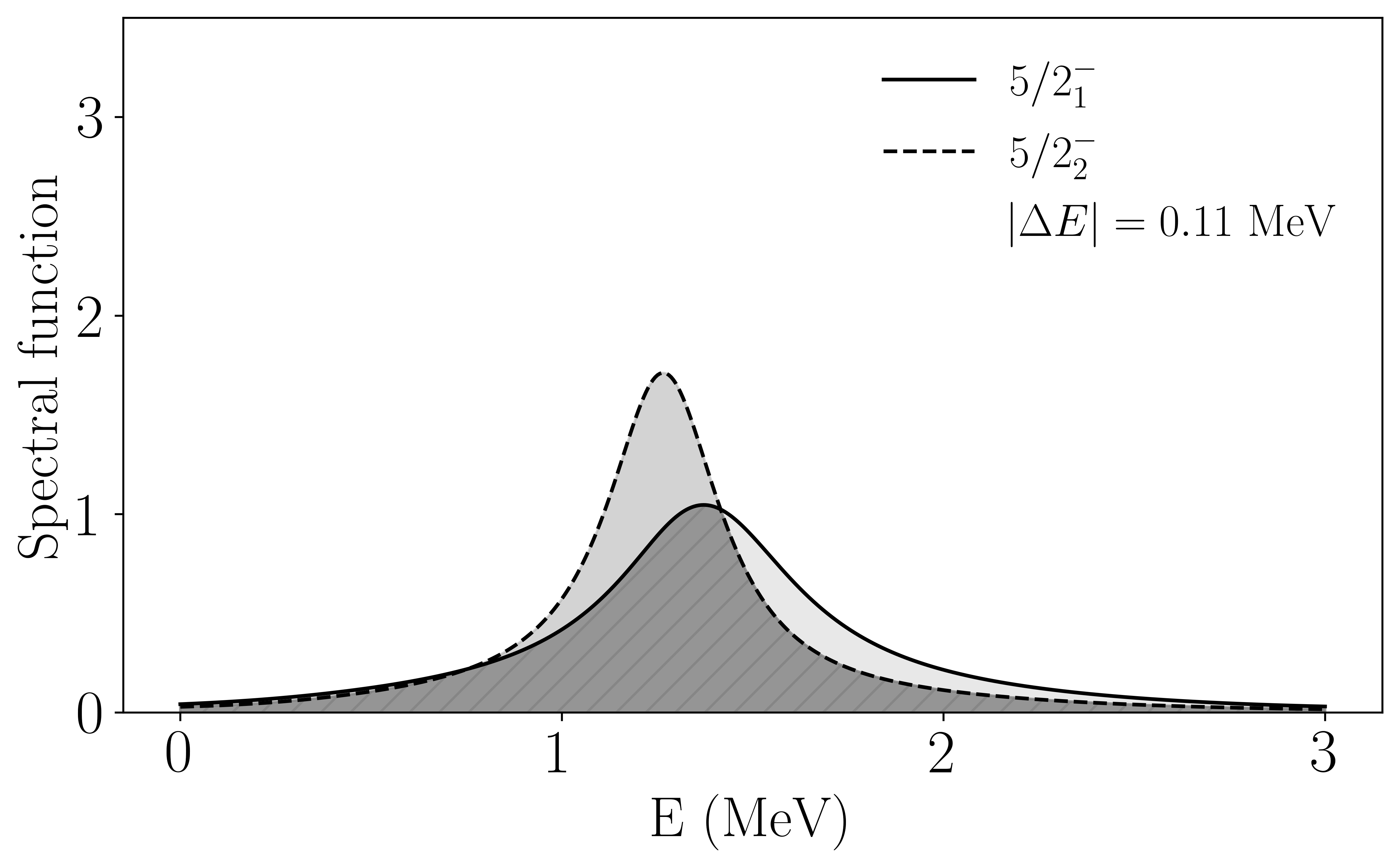}
        \put(12, 55){\large\textbf{c)}}
    \end{overpic}
    \caption{Reduced transitions probabilities B(E2;$5/2^-_{1,2}\rightarrow 3/2^-_{\mathrm{g.s.}}$) in $^7$Be near an exceptional point. Panel (a) shows $\mathrm{Re}[B(E2)]$ for each doublet member and their sum versus the energy separation $\Delta E= E_{5/2^-_1}-E_{5/2^-_2}$. Panels (b) and (c) show the spectral functions for $|\Delta E|=0.44$ MeV and $0.11$ MeV. The EP produces non-physical values of the individual transition probabilities, while their sum remains physical.}
    \label{spf_7Be}
\end{figure}

Panel (a) in Figure~\ref{spf_7Be} shows how the reduced transition probabilities B(E2;$5/2^-_{1,2}\xrightarrow{}3/2^-_{\mathrm{g.s}}$) of $^7$Be evolves as a function of the energy separation between the states in the doublet, defined as $\Delta E= E_{5/2^-_1}-E_{5/2^-_2}$. Panels (b) and (c) are the spectral functions of the states for two different separation energies and illustrate the degree of overlap between the two resonances. One can see that it is not necessary to be exactly at the EP to observe its effects. In this case, even at a separation of $|\Delta E|=0.44$ MeV, an imprint of the EP manifests itself as the appearance of non-physical values of the transition probability coming from the loss of individuality of each resonance mediated by the EP. Just as in the previous section, the sum of both contributions from both states remains smooth, physical and continuous throughout the whole evolution, solidifying the fact that these two states should not be taken as independent from each other, but as a singular entity. 

\begin{figure}
    \centering
    \begin{overpic}[width=0.95\linewidth]{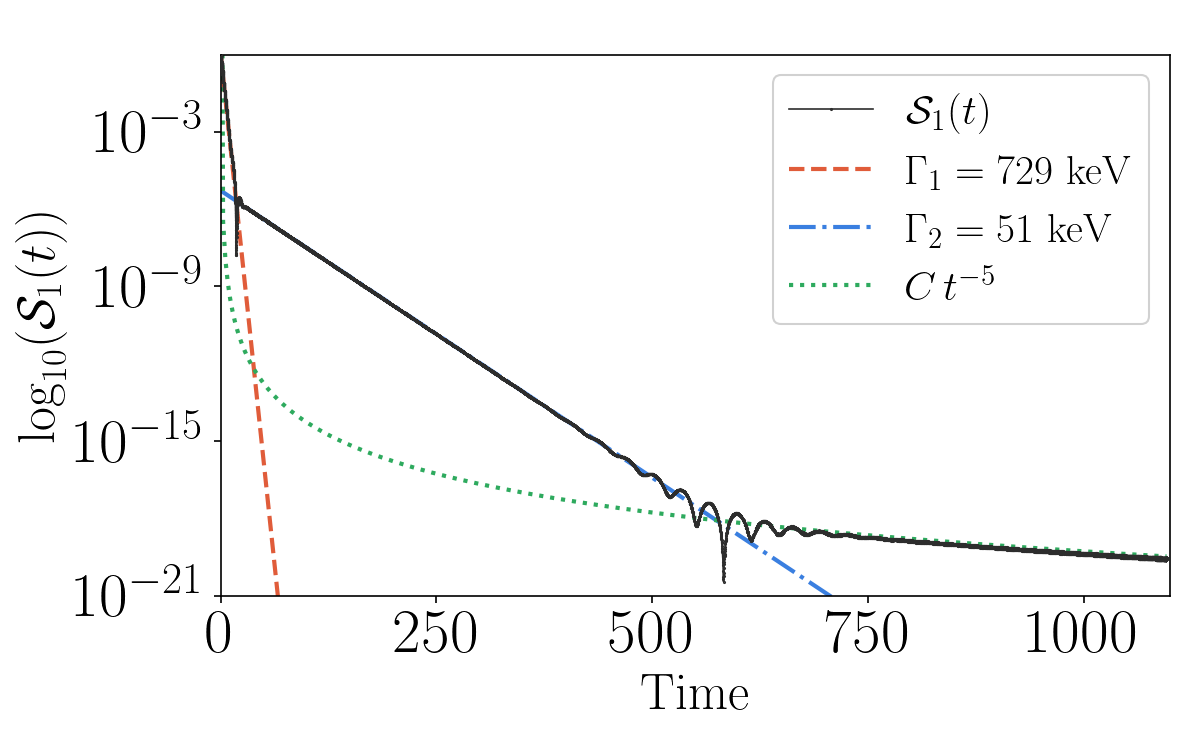}
        \put(22, 52){\large\textbf{a)}}
    \end{overpic}\\
    \begin{overpic}[width=0.95\linewidth]{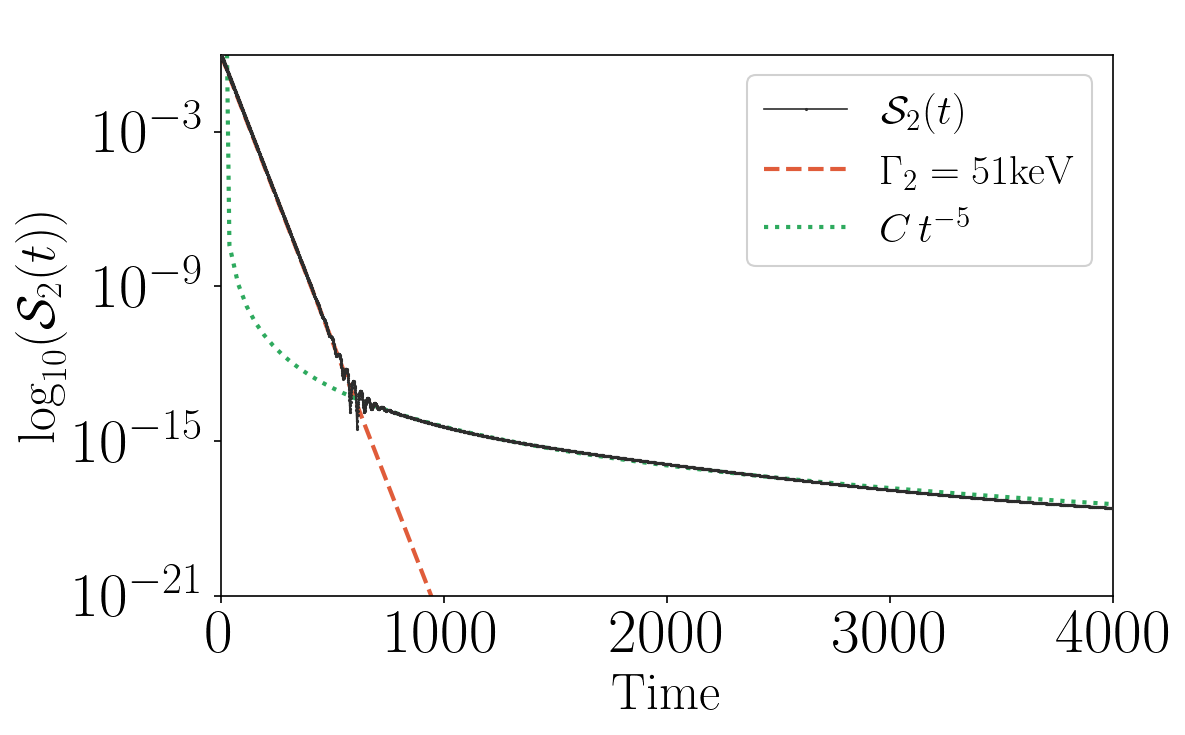}
        \put(22, 52){\large\textbf{b)}}
    \end{overpic}\\
    \begin{overpic}[width=0.95\linewidth]{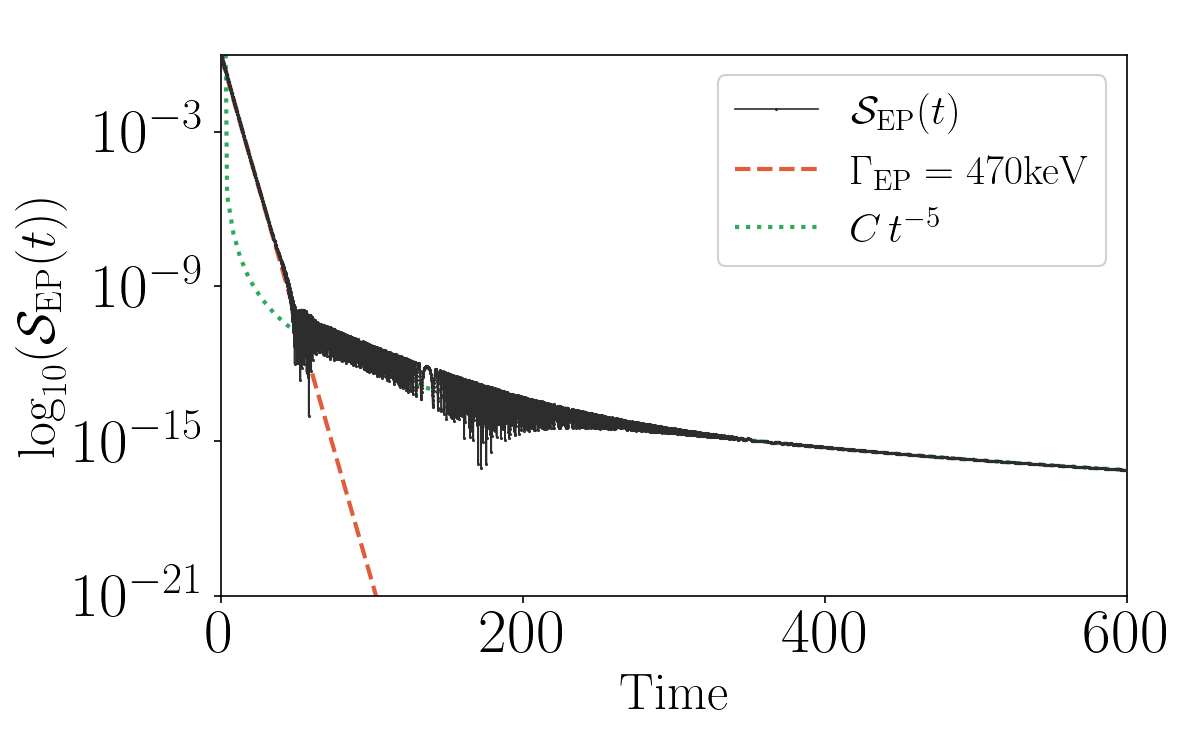}
        \put(22, 52){\large\textbf{c)}}
    \end{overpic}\\
    \begin{overpic}[width=0.95\linewidth]{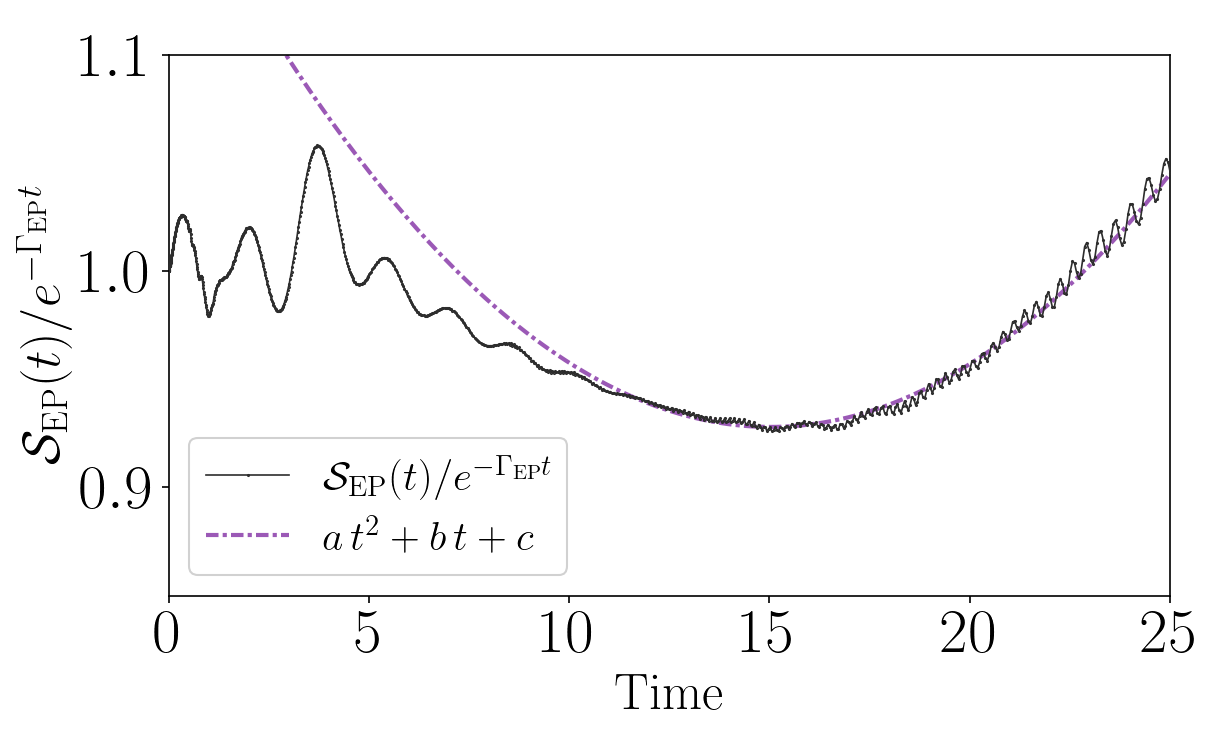}
        \put(16, 50){\large\textbf{d)}}
    \end{overpic}
    \caption{ Panels (a) and (b): Survival probabilities of the $5/2^-_1$  and $5/2^-_2$ resonances at the starting point. Panel (c): Survival probabilities of the merged $5/2^-_{\rm EP}$ state at the exceptional point. Panel (d) shows the 
intermediate region of the EP survival probability divided by $\exp(-\Gamma_{\rm EP} t)$ with $\Gamma_{\rm EP} = 476$ keV, together with a parabolic fit of the form $a + bt + ct^2$.}
    \label{sp_7Li}
\end{figure}

Figure~\ref{sp_7Li} shows the survival probabilities for the $5/2^-$ doublet at the starting point and the EP. In a) and b), in a similar fashion to the $^6\text{Li}$ case, one can see the three expected regimes, with the lower state showing two exponentially decaying regions. The power law region behavior remains proportional to $t^{-5}$ due to the composition of the wave function being mainly $p$-wave. In c), the exponential region appears to decay with a coefficient that differs slightly from $\Gamma_{\text{EP}}=476$ keV. In panel (d), proceeding in the same manner as in the previous section, explains this difference; this region also behaves as an exponential modulated by a quadratic function. Additionally, one can observe the oscillations in the survival probability at short times, arising from the interference between different energy components of the initial state. Each energy eigenstate contributes a phase $e^{-iEt/\hbar}$, and when multiple components are present, they periodically rephase to partially restore the initial state before the irreversible exponential decay takes over. The oscillatory pattern seen on top of the parabola comes from the transitory region in-between the regimes. 

\begin{figure}
    \centering
    \begin{overpic}[width=0.95\linewidth]{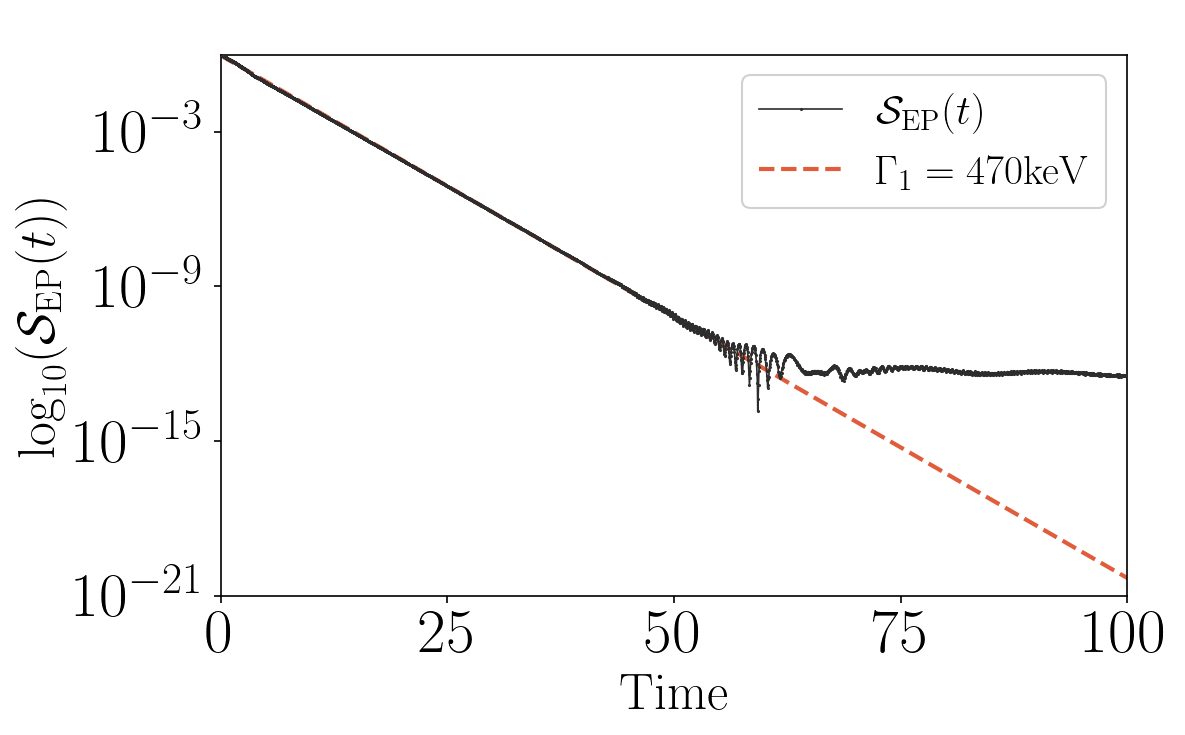}
        \put(20, 49){\large\textbf{a)}}
    \end{overpic}\\
    \begin{overpic}[width=0.95\linewidth]{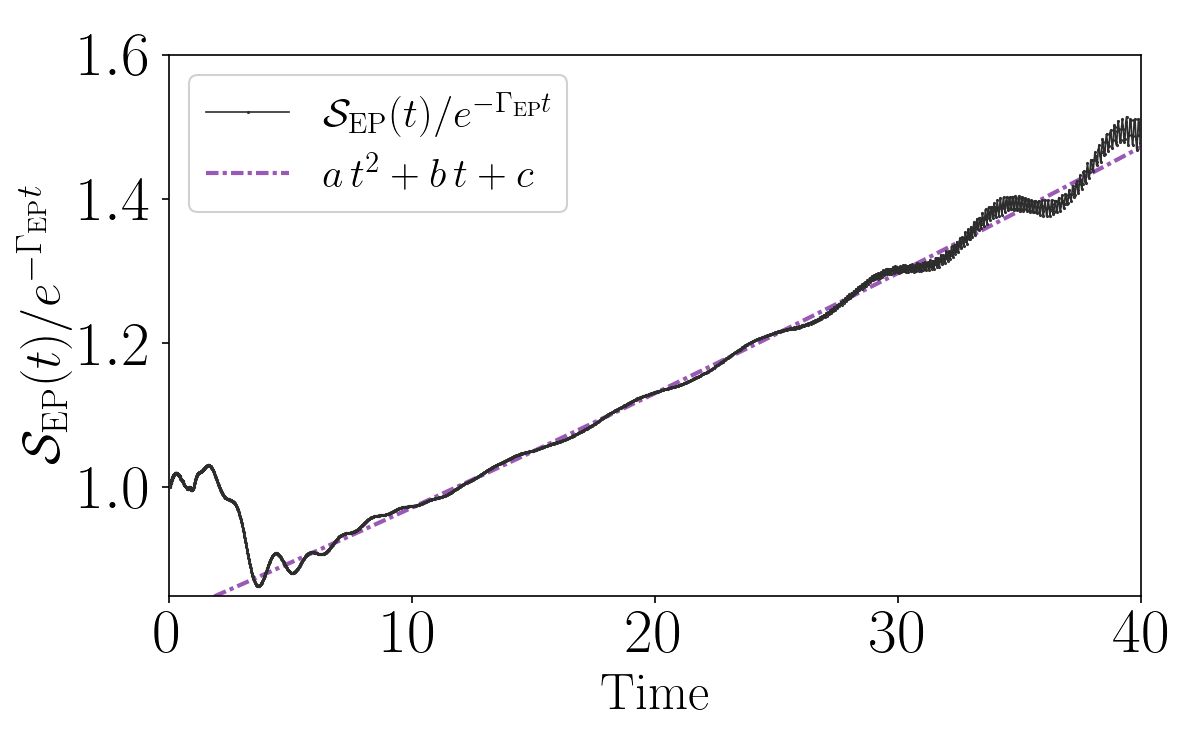}
        \put(88, 52){\large\textbf{b)}}
    \end{overpic}
    \caption{ Panel (a): Exponential region of the survival probability of the $^7\text{Be}$ EP 
state. Panel (b): The same data divided by $\exp(-\Gamma_{\rm EP} t)$ with 
$\Gamma_{\rm EP} = 477$ keV. The result is modulated by a 
polynomial in $t$ with a negligible quadratic term, plus residual 
oscillations from the transitory region.}
    \label{sp_7Be}
\end{figure}

Figure~\ref{sp_7Be} shows the exponential region of the EP state in $^7\text{Be}$. In panel (a) one can see, as in $^7\text{Li}$, that an exponential decay fits fairly well, but the decay coefficient does not match with the width of the state of $\Gamma_{\text{EP}}=477$ keV. In panel (b), after dividing the result by a pure exponential as in the previous cases, it is possible to see that this region is also modulated by a polynomial, though with a negligible quadratic term and an oscillatory pattern on top coming in from the transitory region. The suppression of the quadratic term with respect to the $^7\text{Li}$ and $^6\text{Li}$ cases may be linked to the presence of a second open decay channel at the EP energy: part of the decay strength could be redistributed into the additional ${}^4\text{He}({}^3\text{He},{}^3\text{He}){}^4\text{He}$ channel, possibly weakening the deviation from pure exponential behavior in the elastic proton channel. This would parallel the channel-dependent attenuation of the split-peak signature observed in the cross sections.

\subsection{\texorpdfstring{$^8$Be}{8Be}} \label{sec:A8}
The nucleus in question was modeled by $^4\text{He}$ core with 4 valence 
nucleons and using a channel basis comprising $[^4\text{He}(0_1^+) \otimes  {}^4\text{H}(L_j)]^{J^\pi}$, $[^6\text{Li}(K_i^\pi) \otimes  {}^2\text{H}(L_j)]^{J^\pi}$,  $[^7\text{Li}(K_i^\pi) \otimes p(l_j)]^{J^\pi}$ and $[^7\text{Be}(K_i^\pi) \otimes n(l_j)]^{J^\pi}$ channels. The cluster channels were constructed by coupling the partial waves $L_j $= ${^1S}_{0}$, ${^1P}_{1}$, ${^1D}_{2}$,${^1F}_{3}$, ${^3G}_{4}$ of the $^4\text{He}$ projectile
\begin{figure}
    \centering
    \includegraphics[width=0.95\linewidth]{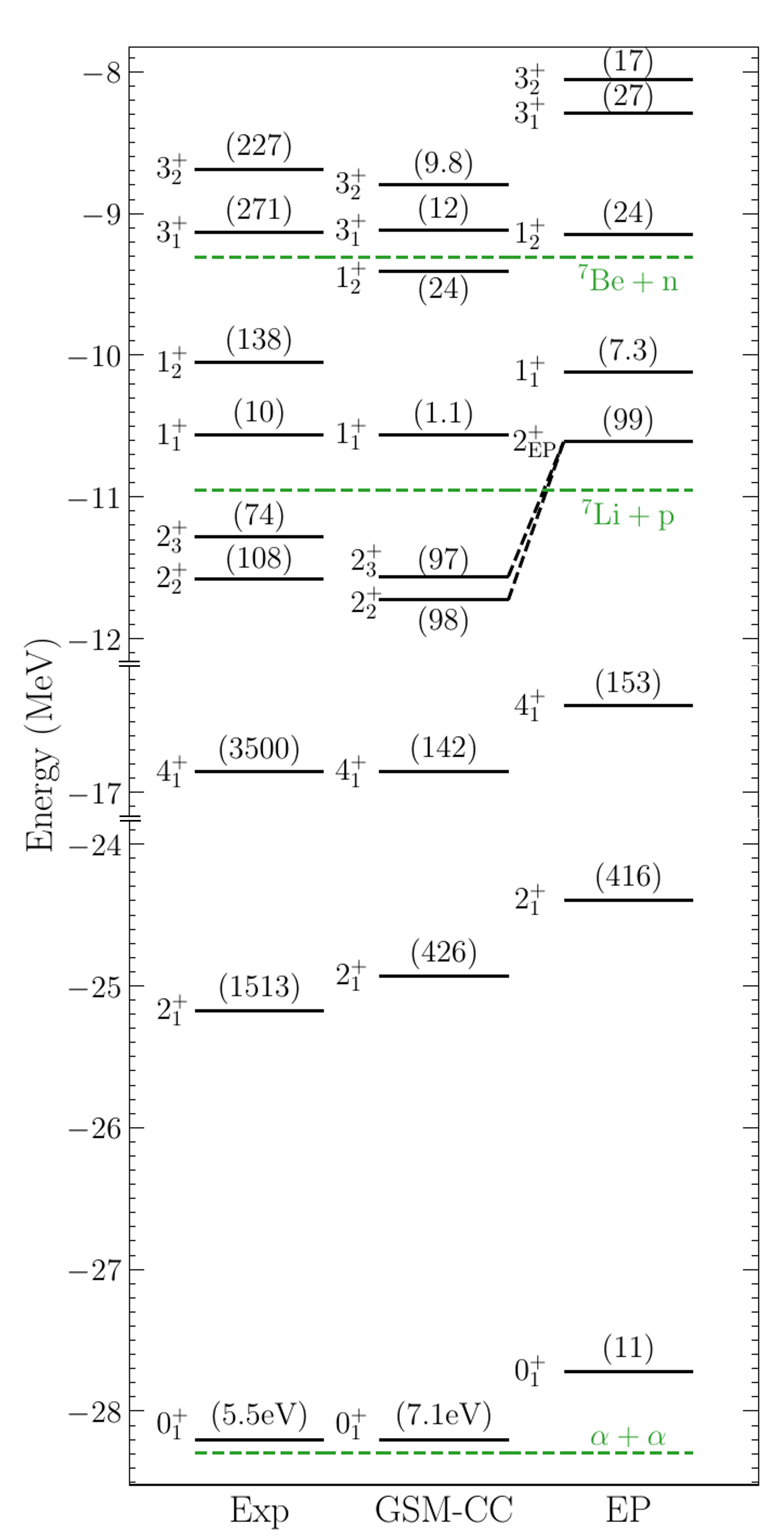}
    \caption{Experimental spectrum of $^8\text{Be}$ is compared with the GSM-CC spectrum.  The widths (in keV) of resonances are given in parentheses.  On the right-hand side, the spectrum for which the two $2^+$ states coalesce.}
    \label{8be_spectrum}
\end{figure}
with the inert core $^4\text{He}$ in the ground state $0^+$ and the partial waves $L_j $= ${^3S}_{1}$, ${^3P}_{0}$, ${^3P}_{1}$, ${^3P}_{2}$, ${^3D}_{1}$, ${^3D}_{2}$, ${^3D}_{3}$ of the wave function of $^2\text{H}$ with the states $K_i^\pi= 1_1^+$, $ 3_1^+$, $ 0_1^+$, $2_1^+$, $2_2^+$ and $1_2^+$ of the $^6\text{Li}$ target. The one-proton and one-neutron channels were built by coupling the partial waves $l_j = s_{1/2}$, $p_{1/2}$, $p_{3/2}$, $d_{3/2}$, $d_{5/2}$, $f_{5/2}$, $f_{7/2}$, $g_{7/2}$, $g_{9/2}$ of their wave functions with the states $K_i^\pi= 3/2_1^-$, $1/2_1^-$, $7/2_1^-$, $5/2_1^-$, $5/2_2^-$, $3/2_2^-$, $1/2_2^-$, $7/2_2^-$  of $^7\text{Li}$ and $^7\text{Be}$ respectively.

The Hamiltonian parameters used for $^8\text{Be}$ are provided in \cite{8be_jose}. The matrix elements for the channel-channel couplings 
involving one-nucleon reaction channels have been re-scaled by corrective factors $c(J^\pi):$ $c(0^+)=1.0151$, $c(4^+)= 1.01225$,  $c(3^+)=0.9992$, $c(2^+)=1.0160 + i0.0016 $. Similarly, for the matrix elements involving the deuteron channels $c_{\rm d}(2^+)=1.0998 -i0.00475$ and $c_{^4\text{He} }(2^+)=1.0998 -i0.00475$ . Figure~\ref{8be_spectrum} shows the obtained spectrum compared to the experimental spectrum. The energy levels are properly reproduced, but the widths are underestimated.

The $2^+$ isospin doublet in $^8$Be near 16.6 and 16.9~MeV excitation has long served as a textbook example of isospin mixing in light nuclei~\cite{Barker1966,Hohloch1967}. Both states decay via $2\alpha$ emission, which is only possible through their $T=0$ component, providing clear experimental evidence for mixing~\cite{Wiringa2013}. This doublet has also attracted attention as a prototypical system of two interacting resonances. Von Brentano analyzed it within a non-Hermitian effective-Hamiltonian framework, showing that mixing through a real off-diagonal interaction produces \emph{energy repulsion} (an avoided crossing in the real part) together with \emph{width attraction} (a convergence of widths)~\cite{vonBrentano1990,vonBrentano1992,vonBrentano1996}. Building on this picture, Hern\'andez and Mondrag\'on proposed that the doublet could be viewed as ``an example of accidental resonance degeneracy'', showing numerically that an exceptional point associated with a double pole of the $S$-matrix can be reached by varying two parameters~\cite{HernandezMondragon1994}. 

\begin{figure}
    \centering
    \includegraphics[width=0.95\linewidth]{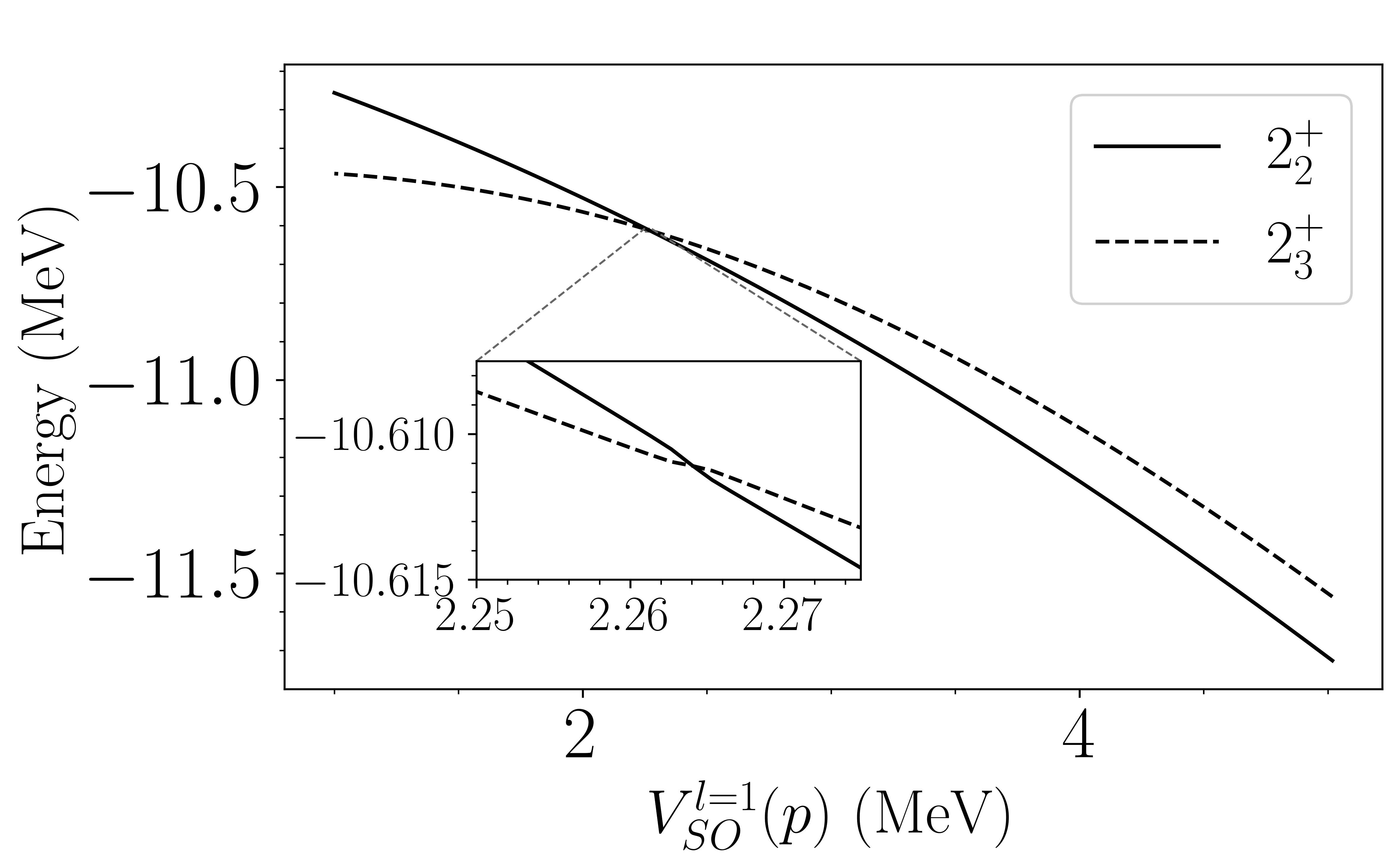}
    \includegraphics[width=0.95\linewidth]{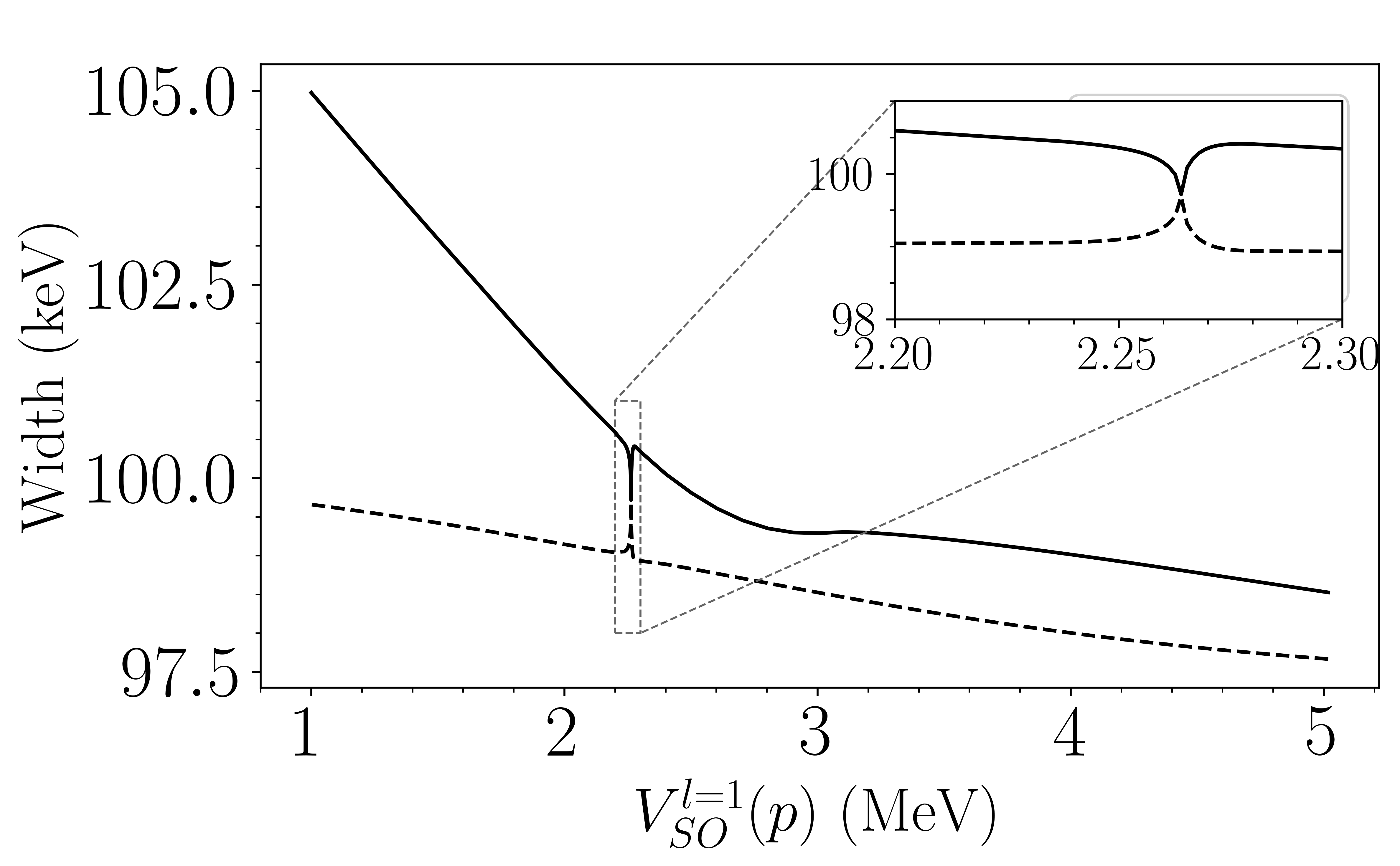}
    \caption{Trajectories of the $2^+$ doublet towards the EP, exhibiting the characteristic square-root behavior of energies and widths.}
        \label{8be_ew}
\end{figure}

\begin{figure}
    \centering
    \includegraphics[width=0.95\linewidth]{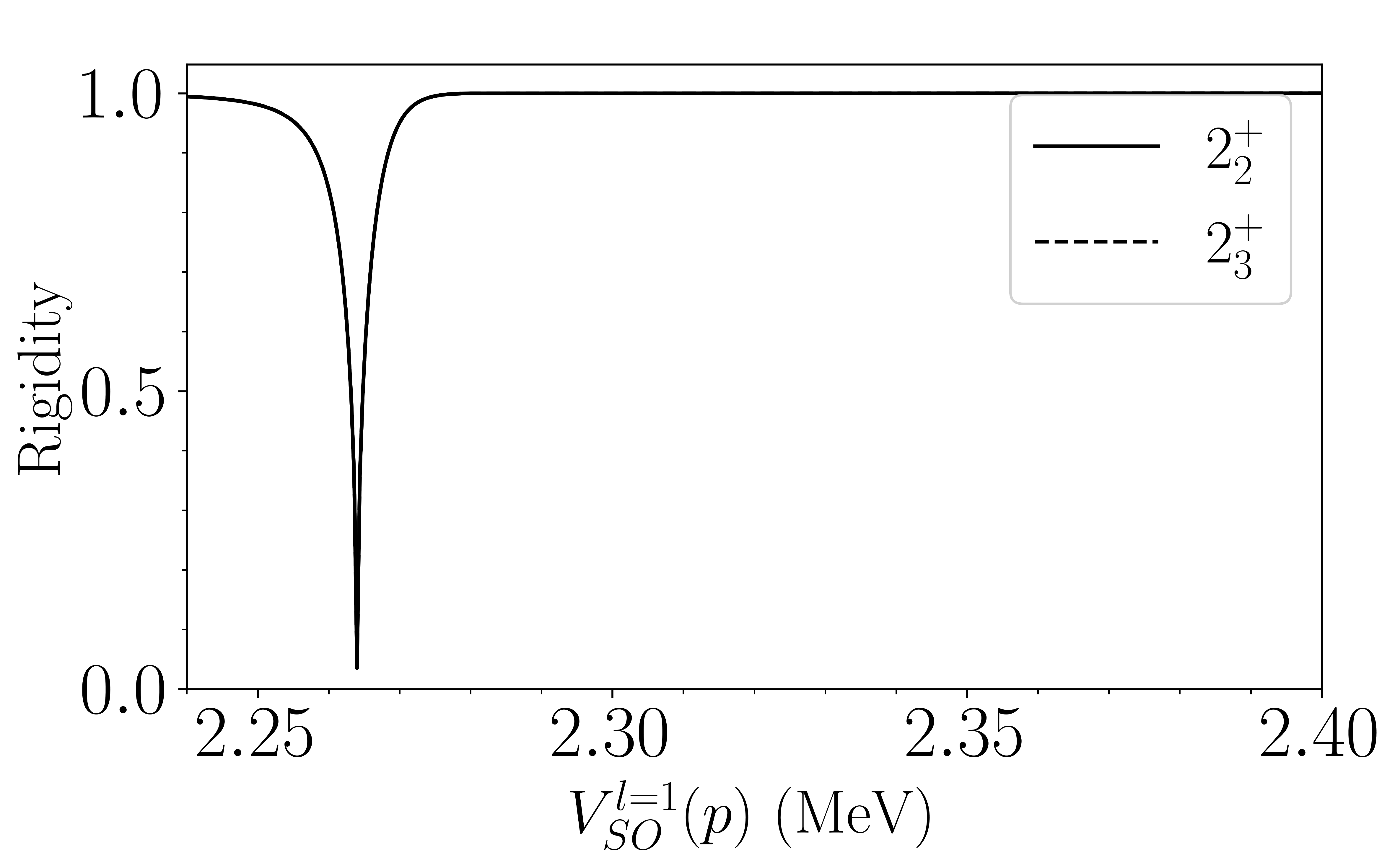}
    \caption{The phase rigidity of $2^+$ resonances is plotted as a function of the control parameters. The phase rigidity becomes equal to zero at the EP.}
    \label{8be_rigidity}
\end{figure}

Similarly to the previous cases, we have chosen the strengths of $\ell=1$ spin-orbit one-body potential for protons and neutrons as the two variable parameters to find the EP. The EP is found at $V^{(l=1)}_{SO}(p)=2.2640$ and $V^{(l=1)}_{SO}(n)=1.9116$. The overall spectrum at the EP is slightly modified from the starting point, although the doublet coalesces at around 1 MeV above their initial position. Figure~\ref{8be_ew} displays the coalescence of the energies and total widths of the $2^+$ states. The rigidity in Figure~\ref{8be_rigidity} shows that the states barely interact with each other, $r=1$, throughout most of the evolution of the parameters, then it decays very sharply in a very narrow range of parameters to $r=0$ signaling the appearance of the EP. Despite this quick decay with respect to the variation of the parameters, the expected square-root behavior of the energies and widths around the EP is still present, as shown in the insets of Figure~\ref{8be_ew}.

\begin{figure}
    \centering
    \includegraphics[width=0.95\linewidth]{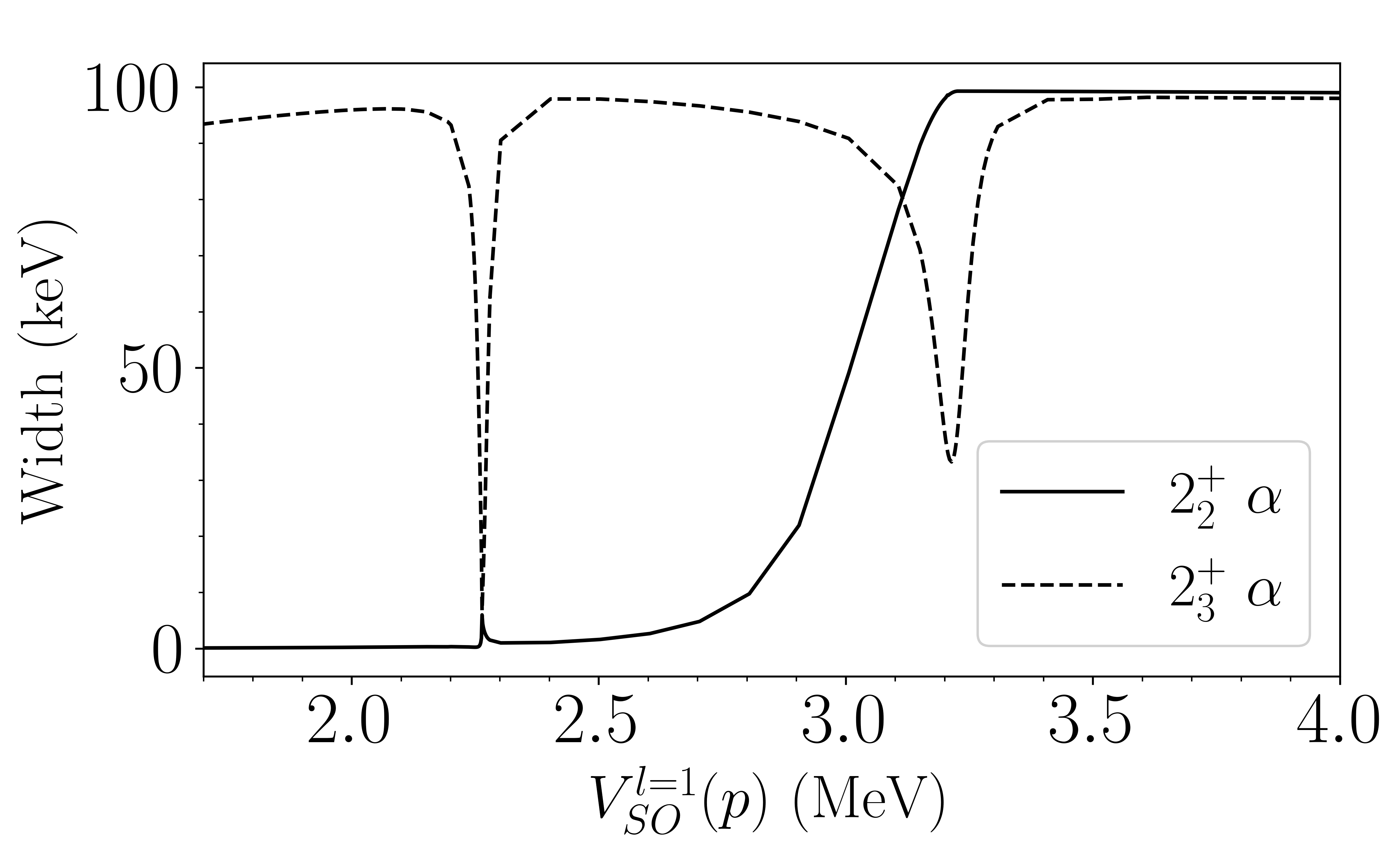}
    \includegraphics[width=0.95\linewidth]{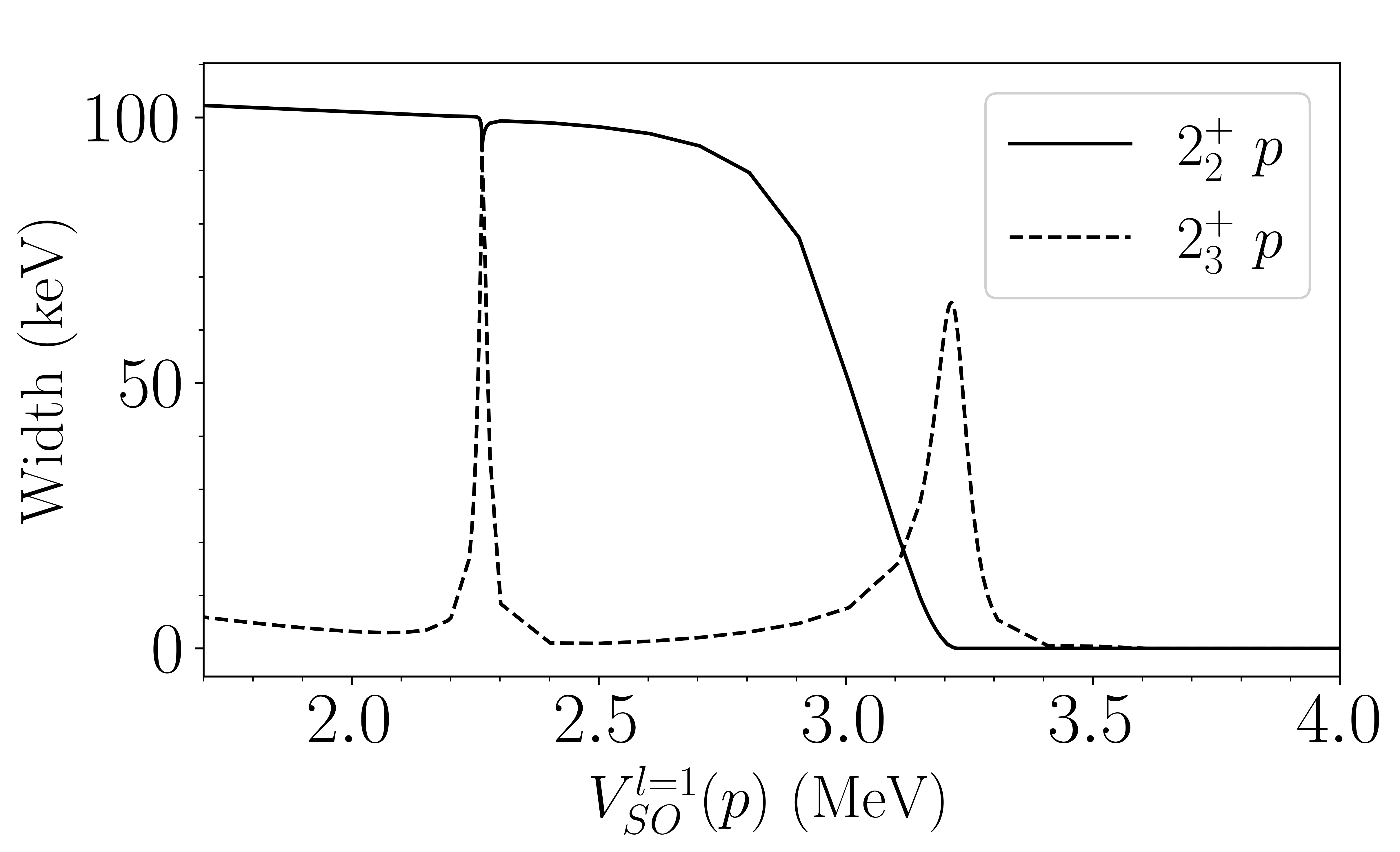}
    \caption{Trajectories of the proton and alpha partial widths towards the EP.}
        \label{8be_pw}
\end{figure}

At the starting point of the trajectory (right side of Figure~\ref{8be_pw}), both states lie below the $^{7}\mathrm{Li}+p$ threshold and carry their entire width in the $\alpha$ channel. This coupling to the $\alpha$ channel is too weak and, hence, not capable of producing the exceptional point. Although the states lie far above the $\alpha+\alpha$ threshold, their $\alpha$ widths are small, reflecting their predominantly nucleonic rather than $\alpha$-clustered structure, so the coupling that $\alpha$ emission provides is weak. Below the proton threshold the system therefore exhibits at most an avoided crossing, with the eigenvectors remaining distinct.

As the spin-orbit parameters are decreased and the states cross the
$^{7}\mathrm{Li}+p$ threshold, a new coupling switches on, supplying the necessary continuum coupling that drives the states to coalesce. The decay structure responds accordingly, and it does so in line with the known structure of the doublet. The lower state is experimentally identified as having a dominant $^{7}\mathrm{Li}+p$ single-particle structure, and it is accordingly this state (the $2^{+}_{2}$) that progressively transfers its width from the $\alpha$ to the proton channel once that channel opens, becoming almost purely proton-decaying by $V^{(l=1)}_{SO}(p)\approx 2.5~\mathrm{MeV}$, while the $2^{+}_{3}$ state briefly loses its $\alpha$ width near $V^{(l=1)}_{SO}(p)\approx 2.3~\mathrm{MeV}$ before recovering it. Since the $\alpha$ width measures the $T=0$ content of a state and the proton width its nucleon-accessible ($T=1$) content, this complementary redistribution is the two states demixing toward good isospin on either side of the crossing: the [$^{7}\mathrm{Li}+p$]-structured state becomes the $T=1$-like (proton-decaying) state and its partner the $T=0$-like ($\alpha$-decaying) state. The resulting maximal channel separation is the most isospin-pure configuration along the trajectory, and it is undone as the open proton channel drives the two states together and they coalesce, exchanging their decay character across the EP.

Despite this dramatic reshuffling of partial widths, the total widths (Figure~\ref{8be_ew}) remain remarkably stable throughout the trajectory, varying by only a few keV around $\sim\!100~\mathrm{keV}$. The bifurcation is therefore almost entirely a redistribution among channels rather than a change in the overall decay rate, consistent with a picture in which the total coupling to the continuum varies smoothly while only its partition between the isospin-selective $\alpha$ channel and the isospin-mixing proton channel changes. The exceptional point thus manifests not as a change in lifetime but as a topological reorganization of the decay channels: a redistribution of the $T=0$ and $T=1$ decay strength, driven by the opening of the proton channel, that collapses the two states into a single coalesced state at the branch point.

\section{Conclusions} \label{sec:conclusions}
We have presented a microscopic study of exceptional points in light nuclei within the coupled-channel Gamow Shell Model. We have shown that special care has to be taken when analyzing overlapping resonances of the same spin and parity, because EP-induced effects, which cannot be accounted for in conventional $R$-matrix and Breit-Wigner approaches, may start to appear even at a considerable distance from the EP.

The four systems studied here illustrate complementary aspects of the same underlying singularity. In $^6\text{Li}$, the EP lies below the proton threshold and the elastic cross section displays the textbook split-peak signature; the spectroscopic factors diverge individually as $1/\sqrt{\varepsilon}$ while their sum 
stays finite, confirming that the two-dimensional subspace spanned by the coalescing states remains physically meaningful even when individual eigenstates do not. In $^7\text{Li}$, the EP is reached with only one open decay channel and lies very close in parameter space to the physical Hamiltonian, in contrast to the more pronounced parameter excursion required in $^6\text{Li}$; this illustrates that closeness of the resonances in energy does not by itself guarantee closeness of the EP in parameter space. Its isospin mirror $^7\text{Be}$ exhibits the same $5/2^-$ doublet, but with the EP located above both the proton and the $^3\text{He}$ thresholds, so that two channels are open at the coalescence. The comparison between the two mirror nuclei thus isolates the effect of having additional open decay channels at the EP: the ratio of partial widths between the two channels in $^7\text{Be}$ dictates which cross section retains the split-peak signature and which is suppressed. In $^8\text{Be}$, the EP is reached only above the $^7\text{Li}+p$ threshold, through a channel-induced bifurcation in which the partial widths reshuffle dramatically while the total width remains nearly constant; the EP thus manifests as a topological reorganization of the decay channels rather than as a change in the overall decay rate. Across all four cases, the survival probability at the EP departs from a pure exponential through a polynomial-modulated intermediate regime, and the spectral function in $^6\text{Li}$ shows the merging of two peaks into a single composite structure with a valley at the EP energy.

These results indicate that light nuclei offer a realistic setting for the study of non-Hermitian degeneracies from their experimental signatures: (i) a split-peak structure in elastic scattering with a minimum at $E_{\rm EP}$; (ii) channel-selective enhancement or suppression of cross sections, as in $^7$Be; (iii) anomalous line shapes in cross sections and spectral functions ;and (iv) large difference in magnitude between transition strengths and related observables of the members of the doublet, such as the case for B(E2) transition probabilities.

 The tuning of the atomic nucleus is limited. Therefore, direct experimental observation of the EP is unlikely, and it is necessary to search for observable consequences of the EP in its vicinity. In this work, we demonstrate that there are several EP signatures that are strong and visible relatively far from the EP and therefore experimentally detectable. Experimental studies of resonance pairs near the EP, at the intersection of nuclear structure and nuclear reaction studies, would allow for a deeper understanding of configuration mixing in the continuum and would also provide insight into the analytical structure of the S matrix and its nontrivial poles.

\begin{acknowledgments}
We thank Jose Pablo Linares Fernandez and Simin Wang for stimulating discussions.
We gratefully acknowledge support from the CNRS/IN2P3 Computing Center (Lyon, France) and the CRIANN (Normandy, France) for providing computing and data-processing resources needed for this work.
This work has been supported by the grants of the National Natural Science Foundation of China Nos. 12347106 and 12575124.
\end{acknowledgments}

\bibliography{bib}

@article{ Heiss2010,
	author = {{Heiss, W. D.} and {Nazmitdinov, R. G.}},
	title = {Resonance scattering and singularities of the scattering function},
	DOI= "10.1140/epjd/e2010-00088-5",
	url= "https://doi.org/10.1140/epjd/e2010-00088-5",
	journal = {Eur. Phys. J. D},
	year = 2010,
	volume = 58,
	number = 1,
	pages = "53-56",
	month = "",
}

@article{10.1063/1.4983809,
    author = {Garmon, Savannah and Ordonez, Gonzalo},
    title = {Characteristic dynamics near two coalescing eigenvalues incorporating continuum threshold effects},
    journal = {J. Math. Phys},
    volume = {58},
    number = {6},
    pages = {062101},
    year = {2017},
    month = {06},
    issn = {0022-2488},
    doi = {10.1063/1.4983809},
    url = {https://doi.org/10.1063/1.4983809},

}

@article{Heiss_2012,
doi = {10.1088/1751-8113/45/44/444016},
url = {https://dx.doi.org/10.1088/1751-8113/45/44/444016},
year = {2012},
month = {oct},
publisher = {IOP Publishing},
volume = {45},
number = {44},
pages = {444016},
author = {Heiss, W. D.},
title = {The physics of exceptional points},
journal = {J. Phys. A}
}

@article{PhysRevLett.103.123003,
  title = {Resonance Coalescence in Molecular Photodissociation},
  author = {Lefebvre, R. and Atabek, O. and \ifmmode \check{S}\else \v{S}\fi{}indelka, M. and Moiseyev, N.},
  journal = {Phys. Rev. Lett.},
  volume = {103},
  issue = {12},
  pages = {123003},
  numpages = {4},
  year = {2009},
  month = {Sep},
  publisher = {American Physical Society},
  doi = {10.1103/PhysRevLett.103.123003},
  url = {https://link.aps.org/doi/10.1103/PhysRevLett.103.123003}
}

@article{Heiss_2005,
doi = {10.1088/0305-4470/38/9/002},
url = {https://dx.doi.org/10.1088/0305-4470/38/9/002},
year = {2005},
month = {feb},
publisher = {},
volume = {38},
number = {9},
pages = {1843},
author = {Heiss, W D and Scholtz, F G and Geyer, H B},
title = {The large N behaviour of the Lipkin model and exceptional points},
journal = {J. Phys. A}
}

@article{PhysRevC.80.034619,
  title = {Exceptional points in the scattering continuum},
  author = {Oko\l{}owicz, J. and P\l{}oszajczak, M.},
  journal = {Phys. Rev. C},
  volume = {80},
  issue = {3},
  pages = {034619},
  numpages = {7},
  year = {2009},
  month = {Sep},
  publisher = {American Physical Society},
  doi = {10.1103/PhysRevC.80.034619},
  url = {https://link.aps.org/doi/10.1103/PhysRevC.80.034619}
}

@article{michel:in2p3-00331689,
  TITLE = {{Shell Model in the Complex Energy Plane}},
  AUTHOR = {Michel, N. and Nazarewicz, W. and Ploszajczak, M. and Vertse, T.},
  URL = {https://in2p3.hal.science/in2p3-00331689},
  NOTE = {Topical Review},
  JOURNAL = {{J. Phys. G}},
  HAL_LOCAL_REFERENCE = {GANIL 09 01 T},
  PUBLISHER = {{IOP Publishing}},
  VOLUME = {36},
  PAGES = {013101},
  YEAR = {2009},
  DOI = {10.1088/0954-3899/36/1/013101},
  HAL_ID = {in2p3-00331689},
  HAL_VERSION = {v1},
}

@article{Wang2017,
  title = {Structure and decays of nuclear three-body systems: The {Gamow} coupled-channel method in $\mathrm{Jacobi}$ coordinates},
  author = {Wang, S. M. and Michel, N. and Nazarewicz, W. and Xu, F. R.},
  journal = {Phys. Rev. C},
  volume = {96},
  issue = {4},
  pages = {044307},
  numpages = {10},
  year = {2017},
  month = {Oct},
  publisher = {American Physical Society},
  doi = {10.1103/PhysRevC.96.044307}
}

@article{PhysRevC.89.034624,
  title = {Gamow shell model description of proton scattering on $^{18}\mathrm{Ne}$},
  author = {Jaganathen, Y. and Michel, N. and P\l{}oszajczak, M.},
  journal = {Phys. Rev. C},
  volume = {89},
  issue = {3},
  pages = {034624},
  numpages = {12},
  year = {2014},
  month = {Mar},
  publisher = {American Physical Society},
  doi = {10.1103/PhysRevC.89.034624},
  url = {https://link.aps.org/doi/10.1103/PhysRevC.89.034624}
}

@article{10.1143/PTP.62.981,
    author = {Furutani, Hiroshi and Horiuchi, Hisashi and Tamagaki, Ryozo},
    title = {{Cluster-Model Study of the $T=1$ States in $A=4$ System: $^3$He+$p$ Scattering}},
    journal = {Prog. Theo. Phys.},
    volume = {62},
    number = {4},
    pages = {981-1002},
    year = {1979},
    month = {10},
    issn = {0033-068X},
    doi = {10.1143/PTP.62.981},
    url = {https://doi.org/10.1143/PTP.62.981},
    eprint = {https://academic.oup.com/ptp/article-pdf/62/4/981/5338838/62-4-981.pdf},
}

@article{PhysRevC.108.044616,
  title = {Description of $^{7}\mathrm{Be}$ and $^{7}\mathrm{Li}$ within the Gamow shell model},
  author = {Fernandez, J. P. Linares and Michel, N. and P\l{}oszajczak, M. and Mercenne, A.},
  journal = {Phys. Rev. C},
  volume = {108},
  issue = {4},
  pages = {044616},
  numpages = {16},
  year = {2023},
  month = {Oct},
  publisher = {American Physical Society},
  doi = {10.1103/PhysRevC.108.044616},
  url = {https://link.aps.org/doi/10.1103/PhysRevC.108.044616}
}

@Inbook{Michel2021,
author="Michel, Nicolas
and P{\l}oszajczak, Marek",
title="The Unification of Structure and Reaction Frameworks",
bookTitle="Gamow Shell Model: The Unified Theory of Nuclear Structure and Reactions",
year="2021",
publisher="Springer International Publishing",
address="Cham",
pages="401--498",
isbn="978-3-030-69356-5",
doi="10.1007/978-3-030-69356-5_9",
url="https://doi.org/10.1007/978-3-030-69356-5_9"
}

@Inbook{Kato1995,
author="Kato, Tosio",
title="Perturbation theory in a finite-dimensional space",
bookTitle="Perturbation Theory for Linear Operators",
year="1995",
publisher="Springer Berlin Heidelberg",
address="Berlin, Heidelberg",
pages="62--126",
isbn="978-3-642-66282-9",
doi="10.1007/978-3-642-66282-9_2",
url="https://doi.org/10.1007/978-3-642-66282-9_2"
}

@inbook{Moiseyev_2011, place={Cambridge}, title={The self-orthogonality phenomenon}, booktitle={Non-Hermitian Quantum Mechanics}, publisher={Cambridge University Press}, author={Moiseyev, Nimrod}, year={2011}, pages={323–374},
isbn="9780511976186",
doi="10.1017/CBO9780511976186",
url="https://doi.org/10.1017/CBO9780511976186"}

@article{Rotter_2015,
doi = {10.1088/0034-4885/78/11/114001},
url = {https://dx.doi.org/10.1088/0034-4885/78/11/114001},
year = {2015},
month = {oct},
publisher = {IOP Publishing},
volume = {78},
number = {11},
pages = {114001},
author = {Rotter, I. and Bird, J. P.},
title = {A review of progress in the physics of open quantum systems: theory and experiment},
journal = {Rep. Prog. Phys.}
}

@misc{ENSDF,
note = "Evaluated Nuclear Structure Data File (ENSDF),
http://www.nndc.bnl.gov/ensdf/"
}

@article{OKOLOWICZ2003271,
title = {Dynamics of quantum systems embedded in a continuum},
journal = {Physics Reports},
volume = {374},
number = {4},
pages = {271-383},
year = {2003},
issn = {0370-1573},
doi = {https://doi.org/10.1016/S0370-1573(02)00366-6},
url = {https://www.sciencedirect.com/science/article/pii/S0370157302003666},
author = {J. Okołowicz and M. Płoszajczak and I. Rotter}
}

@article{cardona,
	author = {{Cardona Ochoa, David} and {Płoszajczak, Marek} and {Michel, Nicolas}},
	title = {Gamow shell model description of exceptional point in 7Li},
	DOI= "10.1051/epjconf/202534201023",
	url= "https://doi.org/10.1051/epjconf/202534201023",
	journal = {EPJ Web Conf.},
	year = 2025,
	volume = 342,
	pages = "01023",}

@article{PhysRevResearch.5.033042,
  title = {Petermann factors and phase rigidities near exceptional points},
  author = {Wiersig, Jan},
  journal = {Phys. Rev. Res.},
  volume = {5},
  issue = {3},
  pages = {033042},
  numpages = {9},
  year = {2023},
  month = {Jul},
  publisher = {American Physical Society},
  doi = {10.1103/PhysRevResearch.5.033042},
  url = {https://link.aps.org/doi/10.1103/PhysRevResearch.5.033042}
}

@article{Heiss_2008,
doi = {10.1088/1751-8113/41/24/244010},
url = {https://doi.org/10.1088/1751-8113/41/24/244010},
year = {2008},
month = {jun},
publisher = {},
volume = {41},
number = {24},
pages = {244010},
author = {Heiss, W D},
title = {Chirality of wavefunctions for three coalescing levels},
journal = {Journal of Physics A: Mathematical and Theoretical}
}

@article{PhysRevB.99.241403,
  title = {Anisotropic exceptional points of arbitrary order},
  author = {Xiao, Yi-Xin and Zhang, Zhao-Qing and Hang, Zhi Hong and Chan, C. T.},
  journal = {Phys. Rev. B},
  volume = {99},
  issue = {24},
  pages = {241403},
  numpages = {6},
  year = {2019},
  month = {Jun},
  publisher = {American Physical Society},
  doi = {10.1103/PhysRevB.99.241403},
  url = {https://link.aps.org/doi/10.1103/PhysRevB.99.241403}
}

@article{PhysRevX.6.021007,
  title = {Emergence, Coalescence, and Topological Properties of Multiple Exceptional Points and Their Experimental Realization},
  author = {Ding, Kun and Ma, Guancong and Xiao, Meng and Zhang, Z. Q. and Chan, C. T.},
  journal = {Phys. Rev. X},
  volume = {6},
  issue = {2},
  pages = {021007},
  numpages = {13},
  year = {2016},
  month = {Apr},
  publisher = {American Physical Society},
  doi = {10.1103/PhysRevX.6.021007},
  url = {https://link.aps.org/doi/10.1103/PhysRevX.6.021007}
}

@article{Jaiswal_2023,
doi = {10.1088/1367-2630/acc1fe},
url = {https://doi.org/10.1088/1367-2630/acc1fe},
year = {2023},
month = {mar},
publisher = {IOP Publishing},
volume = {25},
number = {3},
pages = {033014},
author = {Jaiswal, Rimika and Banerjee, Ayan and Narayan, Awadhesh},
title = {Characterizing and tuning exceptional points using Newton polygons},
journal = {New Journal of Physics},
}

@article{PhysRevA.84.013419,
  title = {Fingerprints of exceptional points in the survival probability of resonances in atomic spectra},
  author = {Cartarius, Holger and Moiseyev, Nimrod},
  journal = {Phys. Rev. A},
  volume = {84},
  issue = {1},
  pages = {013419},
  numpages = {5},
  year = {2011},
  month = {Jul},
  publisher = {American Physical Society},
  doi = {10.1103/PhysRevA.84.013419},
  url = {https://link.aps.org/doi/10.1103/PhysRevA.84.013419}
}

@article{PhysRevE.75.027201,
  title = {Rabi oscillations at exceptional points in microwave billiards},
  author = {Dietz, B. and Friedrich, T. and Metz, J. and Miski-Oglu, M. and Richter, A. and Sch\"afer, F. and Stafford, C. A.},
  journal = {Phys. Rev. E},
  volume = {75},
  issue = {2},
  pages = {027201},
  numpages = {4},
  year = {2007},
  month = {Feb},
  publisher = {American Physical Society},
  doi = {10.1103/PhysRevE.75.027201},
  url = {https://link.aps.org/doi/10.1103/PhysRevE.75.027201}
}

@article{LFonda_1978,
doi = {10.1088/0034-4885/41/4/003},
url = {https://doi.org/10.1088/0034-4885/41/4/003},
year = {1978},
month = {apr},
publisher = {},
volume = {41},
number = {4},
pages = {587},
author = {L Fonda and G C Ghirardi and A Rimini},
title = {Decay theory of unstable quantum systems},
journal = {Reports on Progress in Physics}
}

@article{Peshkin_2014,
doi = {10.1209/0295-5075/107/40001},
url = {https://doi.org/10.1209/0295-5075/107/40001},
year = {2014},
month = {aug},
publisher = {EDP Sciences, IOP Publishing and Società Italiana di Fisica},
volume = {107},
number = {4},
pages = {40001},
author = {Peshkin, Murray and Volya, Alexander and Zelevinsky, Vladimir},
title = {Non-exponential and oscillatory decays in quantum mechanics},
journal = {Europhysics Letters}
}

@article{PhysRevResearch.5.023183,
  title = {Probing the nonexponential decay regime in open quantum systems},
  author = {Wang, S. M. and Nazarewicz, W. and Volya, A. and Ma, Y. G.},
  journal = {Phys. Rev. Res.},
  volume = {5},
  issue = {2},
  pages = {023183},
  numpages = {10},
  year = {2023},
  month = {Jun},
  publisher = {American Physical Society},
  doi = {10.1103/PhysRevResearch.5.023183},
  url = {https://link.aps.org/doi/10.1103/PhysRevResearch.5.023183}
}

@article{PhysRevLett.99.173003,
  title = {Exceptional Points in Atomic Spectra},
  author = {Cartarius, Holger and Main, J\"org and Wunner, G\"unter},
  journal = {Phys. Rev. Lett.},
  volume = {99},
  issue = {17},
  pages = {173003},
  numpages = {4},
  year = {2007},
  month = {Oct},
  publisher = {American Physical Society},
  doi = {10.1103/PhysRevLett.99.173003},
  url = {https://link.aps.org/doi/10.1103/PhysRevLett.99.173003}
}

@article{PhysRevLett.112.203901,
  title = {Enhancing the Sensitivity of Frequency and Energy Splitting Detection by Using Exceptional Points: Application to Microcavity Sensors for Single-Particle Detection},
  author = {Wiersig, Jan},
  journal = {Phys. Rev. Lett.},
  volume = {112},
  issue = {20},
  pages = {203901},
  numpages = {5},
  year = {2014},
  month = {May},
  publisher = {American Physical Society},
  doi = {10.1103/PhysRevLett.112.203901},
  url = {https://link.aps.org/doi/10.1103/PhysRevLett.112.203901}
}

@article{PhysRevA.95.022117,
  title = {Resonances in open quantum systems},
  author = {Eleuch, Hichem and Rotter, Ingrid},
  journal = {Phys. Rev. A},
  volume = {95},
  issue = {2},
  pages = {022117},
  numpages = {12},
  year = {2017},
  month = {Feb},
  publisher = {American Physical Society},
  doi = {10.1103/PhysRevA.95.022117},
  url = {https://link.aps.org/doi/10.1103/PhysRevA.95.022117}
}

@article{
doi:10.1126/science.aar7709,
author = {Mohammad-Ali Miri  and Andrea Alù },
title = {Exceptional points in optics and photonics},
journal = {Science},
volume = {363},
number = {6422},
pages = {eaar7709},
year = {2019},
doi = {10.1126/science.aar7709},
URL = {https://www.science.org/doi/abs/10.1126/science.aar7709},
eprint = {https://www.science.org/doi/pdf/10.1126/science.aar7709}}

@article{EHernandez_2000,
doi = {10.1088/0305-4470/33/24/308},
url = {https://doi.org/10.1088/0305-4470/33/24/308},
year = {2000},
month = {jun},
publisher = {},
volume = {33},
number = {24},
pages = {4507},
author = {E Hernández and A Jáuregui and A Mondragón},
title = {Degeneracy of resonances in a double barrier potential},
journal = {Journal of Physics A: Mathematical and General}
}

@article{ 8be_jose,
	author = {{Linares Fern\'andez, Jos\'e Pablo} and {Michel, Nicolas} and {Ploszajczak, Marek}},
	title = {Clusterization in nuclear states at the edge of stability},
	DOI= "10.1051/epjconf/202431100016",
	url= "https://doi.org/10.1051/epjconf/202431100016",
	journal = {EPJ Web Conf.},
	year = 2024,
	volume = 311,
	pages = "00016",
}

@article{CardonaOchoa2026,
  author    = {Cardona Ochoa, David and P{\l}oszajczak, Marek and Michel, Nicolas and Wang, Simin},
  title     = {Double Pole {S}-Matrix Singularity in the Continuum of $^{7}${Be}},
  journal   = {Acta Physica Polonica B Proceedings Supplement},
  volume    = {19},
  pages     = {1-A1},
  year      = {2026},
  doi       = {10.5506/APhysPolBSupp.19.1-A1},
}

@article{Ozdemir2019,
  author  = {\"Ozdemir, {\c{S}}. K. and Rotter, S. and Nori, F. and Yang, L.},
  title   = {Parity--time symmetry and exceptional points in photonics},
  journal = {Nature Materials},
  year    = {2019},
  volume  = {18},
  number  = {8},
  pages   = {783--798},
  doi     = {10.1038/s41563-019-0304-9},
  url     = {https://doi.org/10.1038/s41563-019-0304-9},
  issn    = {1476-4660}
}

@article{PhysRevA.43.4159,
  title = {Transitional regions of finite Fermi systems and quantum chaos},
  author = {Heiss, W. D. and Sannino, A. L.},
  journal = {Phys. Rev. A},
  volume = {43},
  issue = {8},
  pages = {4159--4166},
  numpages = {0},
  year = {1991},
  month = {Apr},
  publisher = {American Physical Society},
  doi = {10.1103/PhysRevA.43.4159},
  url = {https://link.aps.org/doi/10.1103/PhysRevA.43.4159}
}

@article{Mailybaev2005,
  author    = {Mailybaev, A. A. and Kirillov, O. N. and Seyranian, A. P.},
  title     = {Geometric phase around exceptional points},
  journal   = {Physical Review A},
  volume    = {72},
  pages     = {014104},
  year      = {2005},
  doi       = {10.1103/PhysRevA.72.014104}
}

@article{Berry2004,
  author    = {Berry, M. V.},
  title     = {Physics of Nonhermitian Degeneracies},
  journal   = {Czechoslovak Journal of Physics},
  volume    = {54},
  pages     = {1039--1047},
  year      = {2004},
  doi       = {10.1023/B:CJOP.0000044002.05657.04}
}

@article{Doppler2016,
  author    = {Doppler, J{\"o}rg and Mailybaev, Alexei A. and B{\"o}hm, Julian
               and Kuhl, Ulrich and Girschik, Adrian and Libisch, Florian
               and Milburn, Thomas J. and Rabl, Peter and Moiseyev, Nimrod
               and Rotter, Stefan},
  title     = {Dynamically encircling an exceptional point for asymmetric mode switching},
  journal   = {Nature},
  volume    = {537},
  pages     = {76--79},
  year      = {2016},
  doi       = {10.1038/nature18605}
}

@article{Xu2016,
  author    = {Xu, H. and Mason, D. and Jiang, Luyao and Harris, J. G. E.},
  title     = {Topological energy transfer in an optomechanical system with exceptional points},
  journal   = {Nature},
  volume    = {537},
  pages     = {80--83},
  year      = {2016},
  doi       = {10.1038/nature18604}
}

@article{Wiersig2014,
  author    = {Wiersig, Jan},
  title     = {Enhancing the Sensitivity of Frequency and Energy Splitting Detection
               by Using Exceptional Points: Application to Microcavity Sensors
               for Single-Particle Detection},
  journal   = {Physical Review Letters},
  volume    = {112},
  pages     = {203901},
  year      = {2014},
  doi       = {10.1103/PhysRevLett.112.203901}
}

@article{Langbein2018,
  author    = {Langbein, Wolfgang},
  title     = {No exceptional precision of exceptional-point sensors},
  journal   = {Physical Review A},
  volume    = {98},
  pages     = {023805},
  year      = {2018},
  doi       = {10.1103/PhysRevA.98.023805}
}

@article{Berry2003,
  author    = {Berry, M. V.},
  title     = {Mode degeneracies and the {Petermann} excess-noise factor for unstable lasers},
  journal   = {Journal of Modern Optics},
  volume    = {50},
  pages     = {63--81},
  year      = {2003},
  doi       = {10.1080/09500340308234532}
}

@article{Rotter2009,
  author    = {Rotter, Ingrid},
  title     = {A non-{H}ermitian {H}amilton operator and the physics of open quantum systems},
  journal   = {Journal of Physics A: Mathematical and Theoretical},
  volume    = {42},
  number    = {15},
  pages     = {153001},
  year      = {2009},
  doi       = {10.1088/1751-8113/42/15/153001}
}

@article{EleuchRotter2013,
  author    = {Eleuch, H. and Rotter, I.},
  title     = {Width bifurcation and dynamical phase transitions in open quantum systems},
  journal   = {Physical Review E},
  volume    = {87},
  pages     = {052136},
  year      = {2013},
  doi       = {10.1103/PhysRevE.87.052136}
}

@article{LaneThomas1958,
  author    = {Lane, A. M. and Thomas, R. G.},
  title     = {$R$-Matrix Theory of Nuclear Reactions},
  journal   = {Reviews of Modern Physics},
  volume    = {30},
  pages     = {257--353},
  year      = {1958},
  doi       = {10.1103/RevModPhys.30.257}
}

@article{DescouvemontBaye2010,
  author    = {Descouvemont, Pierre and Baye, Daniel},
  title     = {The $R$-matrix theory},
  journal   = {Reports on Progress in Physics},
  volume    = {73},
  number    = {3},
  pages     = {036301},
  year      = {2010},
  doi       = {10.1088/0034-4885/73/3/036301}
}

@book{Satchler1983,
  author    = {Satchler, G. R.},
  title     = {Direct Nuclear Reactions},
  publisher = {Oxford University Press},
  address   = {Oxford},
  year      = {1983}
}

@article{Heiss2004,
  author    = {Heiss, W. D.},
  title     = {Exceptional points of non-{H}ermitian operators},
  journal   = {Journal of Physics A: Mathematical and General},
  volume    = {37},
  number    = {6},
  pages     = {2455--2464},
  year      = {2004},
  doi       = {10.1088/0305-4470/37/6/034}
}

@article{AshidaGongUeda2020,
  author    = {Ashida, Yuto and Gong, Zongping and Ueda, Masahito},
  title     = {Non-{H}ermitian physics},
  journal   = {Advances in Physics},
  volume    = {69},
  number    = {3},
  pages     = {249--435},
  year      = {2020},
  doi       = {10.1080/00018732.2021.1876991}
}

@article{Eleuch2018,
  author    = {Eleuch, Hichem and Rotter, Ingrid},
  title     = {Critical points in two-channel quantum systems},
  journal   = {The European Physical Journal D},
  volume    = {72},
  pages     = {138},
  year      = {2018},
  doi       = {10.1140/epjd/e2018-90031-1},
  publisher = {Springer}
}

@book{Kato1966,
  author    = {Kato, Tosio},
  title     = {Perturbation Theory for Linear Operators},
  publisher = {Springer},
  address   = {Berlin},
  year      = {1966}
}

@article{Heiss1999,
  author  = {Heiss, W. D.},
  title   = {Phases of wave functions and level repulsion},
  journal = {Eur. Phys. J. D},
  volume  = {7},
  pages   = {1--4},
  year    = {1999},
  doi     = {10.1007/s100530050339}
}

@article{Dembowski2001,
  author  = {Dembowski, C. and Gr{\"a}f, H.-D. and Harney, H. L. and Heine, A. and Heiss, W. D. and Rehfeld, H. and Richter, A.},
  title   = {Experimental observation of the topological structure of exceptional points},
  journal = {Phys. Rev. Lett.},
  volume  = {86},
  pages   = {787--790},
  year    = {2001},
  doi     = {10.1103/PhysRevLett.86.787}
}

@article{Brody2014,
  author  = {Brody, Dorje C.},
  title   = {Biorthogonal quantum mechanics},
  journal = {J. Phys. A: Math. Theor.},
  volume  = {47},
  pages   = {035305},
  year    = {2014},
  doi     = {10.1088/1751-8113/47/3/035305}
}

@article{PhysRevC.99.044606,
  title = {Gamow shell model description of $^{4}\mathrm{He}(d,d)$ elastic scattering reactions},
  author = {Mercenne, A. and Michel, N. and P\l{}oszajczak, M.},
  journal = {Phys. Rev. C},
  volume = {99},
  issue = {4},
  pages = {044606},
  numpages = {13},
  year = {2019},
  month = {Apr},
  publisher = {American Physical Society},
  doi = {10.1103/PhysRevC.99.044606},
  url = {https://link.aps.org/doi/10.1103/PhysRevC.99.044606}
}

@article{Dong2022,
  title = {Gamow shell model description of the radiative capture reaction $^{8}\mathrm{Li}(n,\gamma)^{9}\mathrm{Li}$},
  author = {Dong, G. X. and Wang, X. B. and Michel, N. and P\l{}oszajczak, M.},
  journal = {Phys. Rev. C},
  volume = {105},
  pages = {064608},
  year = {2022},
  doi = {10.1103/PhysRevC.105.064608}
}

@article{Chen2025,
  title = {Gamow shell model study of the $^{17}$Ne$(p,p)$ reaction and of isospin symmetry breaking in $^{18}$Na},
  author = {Chen, N. and Li, J. G. and Li, K. H. and Michel, N. and Wang, P. Y. and Zuo, W.},
  journal = {Phys. Rev. C},
  year = {2025},
  note = {arXiv:2510.07693}
}

@article{Berry1984,
  author = {Berry, M. V.},
  title = {Quantal phase factors accompanying adiabatic changes},
  journal = {Proc. R. Soc. Lond. A},
  volume = {392},
  pages = {45--57},
  year = {1984},
  doi = {10.1098/rspa.1984.0023}
}

@article{Simon1983,
  author = {Simon, Barry},
  title = {Holonomy, the Quantum Adiabatic Theorem, and Berry's Phase},
  journal = {Phys. Rev. Lett.},
  volume = {51},
  pages = {2167--2170},
  year = {1983},
  doi = {10.1103/PhysRevLett.51.2167}
}

@article{MisraSudarshan1977,
  author = {Misra, B. and Sudarshan, E. C. G.},
  title = {The Zeno's paradox in quantum theory},
  journal = {J. Math. Phys.},
  volume = {18},
  pages = {756--763},
  year = {1977},
  doi = {10.1063/1.523304}
}

@article{Itano1990,
  author = {Itano, W. M. and Heinzen, D. J. and Bollinger, J. J. and Wineland, D. J.},
  title = {Quantum Zeno effect},
  journal = {Phys. Rev. A},
  volume = {41},
  pages = {2295--2300},
  year = {1990},
  doi = {10.1103/PhysRevA.41.2295}
}

@article{Barker1966,
  author  = {Barker, F. C.},
  title   = {Intermediate coupling shell-model calculations for light nuclei},
  journal = {Nucl. Phys.},
  volume  = {83},
  pages   = {418--448},
  year    = {1966},
  doi     = {10.1016/0029-5582(66)90582-7}
}

@article{Hohloch1967,
  author  = {Hohloch, E. and Wildermuth, K.},
  title   = {Isobaric spin mixing in the $^8$Be 16.62 and 16.92 MeV states},
  journal = {Phys. Lett. B},
  volume  = {24},
  pages   = {121--123},
  year    = {1967},
  doi     = {10.1016/0370-2693(67)90496-0}
}

@article{Wiringa2013,
  author  = {Wiringa, R. B. and Pastore, S. and Pieper, Steven C. and Miller, Gerald A.},
  title   = {Charge-symmetry breaking forces and isospin mixing in $^8$Be},
  journal = {Phys. Rev. C},
  volume  = {88},
  pages   = {044333},
  year    = {2013},
  doi     = {10.1103/PhysRevC.88.044333}
}

@article{vonBrentano1990,
  author  = {von Brentano, P.},
  title   = {Energy repulsion and width attraction in a system of two interacting resonances in $^8$Be},
  journal = {Phys. Lett. B},
  volume  = {246},
  pages   = {320--324},
  year    = {1990},
  doi     = {10.1016/0370-2693(90)90606-7}
}

@article{vonBrentano1992,
  author  = {von Brentano, P.},
  title   = {Energy attraction and width repulsion due to external mixing and an application to $^8$Be},
  journal = {Nucl. Phys. A},
  volume  = {550},
  pages   = {143--149},
  year    = {1992},
  doi     = {10.1016/0375-9474(92)91135-C}
}

@article{vonBrentano1996,
  author  = {von Brentano, P.},
  title   = {On the mixing of two bound and unbound levels: Energy repulsion and width attraction},
  journal = {Phys. Rep.},
  volume  = {264},
  pages   = {57--66},
  year    = {1996},
  doi     = {10.1016/0370-1573(95)00027-5}
}

@article{HernandezMondragon1994,
  author  = {Hern{\'a}ndez, E. and Mondrag{\'o}n, A.},
  title   = {The $2^+$ doublet in $^8$Be: an example of accidental resonance degeneracy},
  journal = {Phys. Lett. B},
  volume  = {326},
  pages   = {1--4},
  year    = {1994},
  doi     = {10.1016/0370-2693(94)91182-7}
}
\bibliographystyle{apsrev4-1}

\end{document}